\documentclass{article}

\usepackage{arxiv}

\usepackage[utf8]{inputenc} 
\usepackage[T1]{fontenc}    
\usepackage{hyperref}       
\usepackage{url}            
\usepackage{booktabs}       
\usepackage{amsfonts}       
\usepackage{nicefrac}       
\usepackage{microtype}      
\usepackage{lipsum}		
\usepackage{graphicx}
\usepackage{natbib}
\usepackage{doi}
\usepackage{multirow}
\usepackage{siunitx}
\usepackage{colortbl}
\usepackage{xcolor}
\usepackage{pgfmath}
\usepackage{graphicx}
\usepackage{caption}
\usepackage{subcaption}
\usepackage{CJKutf8}
\usepackage{cleveref}

\title{The Honorific Effect: Exploring the Impact of Japanese Linguistic Formalities on AI-Generated Physics Explanations}

\date{} 					

\author{ 
        Keisuke Sato \\
	National Institute of Technology, Ibaraki College\\
        Hitachinaka, Ibaraki, Japan\\
	\texttt{skeisuke@ibaraki-ct.ac.jp} \\
}

\renewcommand{\headeright}{Technical Report}
\renewcommand{\shorttitle}{\textit{The Honorific Effect: Exploring the Impact of Japanese Linguistic} }

\hypersetup{
pdftitle={Exploring the Educational Landscape of AI},
pdfsubject={physics.ed-ph, cs.CY, cs.AI},
pdfauthor={Keisuke Sato},
pdfkeywords={Physics Education, Artificial Intelligence, Large Language Models, Conservation of Momentum, Educational Technology, AI-assisted Learning, STEM Education},
}

\begin{document}
\maketitle

\begin{abstract}
This study investigates the influence of Japanese honorifics on the responses of large language models (LLMs) when explaining the law of conservation of momentum. We analyzed the outputs of six state-of-the-art AI models, including variations of ChatGPT, Coral, and Gemini, using 14 different honorific forms. Our findings reveal that honorifics significantly affect the quality, consistency, and formality of AI-generated responses, demonstrating LLMs' ability to interpret and adapt to social context cues embedded in language. Notable variations were observed across different models, with some emphasizing historical context and derivations, while others focused on intuitive explanations. The study highlights the potential for using honorifics to adjust the depth and complexity of AI-generated explanations in educational contexts. Furthermore, the responsiveness of AI models to cultural linguistic elements underscores the importance of considering cultural factors in AI development for educational applications. These results open new avenues for research in AI-assisted education and cultural adaptation in AI systems, with significant implications for personalizing learning experiences and developing culturally sensitive AI tools for global education.
\end{abstract}


\section{Introduction}

With the rapid advancement of artificial intelligence (AI) technologies, their potential applications in education have gained significant attention. Large language models (LLMs), a subset of AI models, hold promise as personalized learning support tools due to their flexible response capabilities and vast knowledge bases. However, the use of AI in education presents several challenges, one of which is the issue of quality and consistency in AI responses \cite{Sarwari2024-yf}. Moreover, the influence of cultural factors on AI responses remains poorly understood. Recent research by Ge et al. \cite{Ge2024-mx} has shown that cultural differences significantly shape people's preferences for and interactions with AI.

The Japanese honorific system is a complex cultural practice that involves choosing appropriate language based on social relationships and context. It is not merely a linguistic feature but a crucial element that significantly impacts the quality and effectiveness of communication. In educational settings, honorifics express relationships between teachers and students, seniors and juniors, and influence the learning environment and knowledge transfer methods. This system provides a unique lens through which to examine the cultural adaptability of AI in educational contexts.

The use of honorifics in interactions with AI can also influence the relationship between humans and AI. For example, the expression "Google Sensei" (Teacher Google) suggests a personification of the search engine and a sense of respect. Similarly, addressing an AI assistant as "Mr./Ms. [Name]" indicates a recognition of the AI as a dialogue partner rather than just a tool. These linguistic choices reflect users' perceptions of AI and may influence the nature of human-AI interactions. Recent research has shown that job titles and perceived roles assigned to AI agents can significantly impact users' perceptions and interactions with these agents, enhancing their likability and trustworthiness \cite{Jeon2022-wv}.

This research aims to bridge the gap between cultural linguistics and AI-assisted education by analyzing how LLM responses vary depending on the use of Japanese honorifics. Specifically, we will compare and analyze the responses of six AI models to the prompt "Please explain the law of conservation of momentum" \cite{sato2024exploringeducationallandscapeai} using 14 different honorifics. This will reveal how honorifics affect the quality, consistency, and complexity of AI explanations.

The significance of this research lies in the following points:
\begin{itemize}
    \item Elucidating the importance of cultural factors in AI-assisted learning: Understanding the impact of honorifics, a linguistic and cultural element, on AI responses can contribute to the development of more effective and culturally appropriate AI educational tools.
    \item Emphasizing the importance of linguistic nuances in AI-human interactions: Analyzing the effects of honorific use on AI responses reveals the importance of subtle linguistic nuances in AI-human communication, providing valuable insights for the design of future AI systems.
    \item Implications for AI literacy education: Understanding how the use of honorifics affects interactions with AI can provide important suggestions for educating students on how to effectively communicate with and utilize AI.
    \item Providing fundamental knowledge for the social implementation of AI technology: Understanding the influence of cultural factors on AI responses reveals important elements to consider in the social implementation of AI technology, contributing to human-centered AI development.
\end{itemize}

This research represents a pioneering effort in demonstrating the importance of cultural factors in AI-human communication, particularly in educational contexts.
Its findings are expected to contribute to the improvement of AI-assisted learning system design, the development of AI literacy education, and the establishment of a richer symbiotic relationship between AI and humans.

\section{Experimental Methods}

\subsection{AI Models}

This study evaluated six state-of-the-art large language models (LLMs), selected for their prominence, diverse approaches to language processing, and potential for educational applications:

\begin{enumerate} 
    \item ChatGPT 3.5 Turbo (G3): A model developed by OpenAI that is widely used for general purposes. 
    \item ChatGPT 4.0 Turbo (G4): An advanced version of the model developed by OpenAI with improved inference capabilities. 
    \item Coral (Command R+) (C): A model developed by Cohere that provides language understanding using a different approach. 
    \item Gemini 1.0 Pro (Ge1p): A multimodal AI model from Google that can handle a variety of different types of input. 
    \item Gemini 1.5 Flash (Ge15f): An optimised version for faster response times. 
    \item Gemini 1.5 Pro (Ge15p): The most advanced model in the Gemini series at the time of writing.  
\end{enumerate}

These models represent a broad spectrum of AI technologies, developed by different companies and research institutions, allowing for a comprehensive assessment of the current state of AI capabilities in educational contexts.

\subsection{Experimental Design}

\subsubsection{Data Collection}

Data collection was conducted from June 29, 2024, to July 3, 2024. Each model was prompted with the following question in Japanese:

\begin{CJK}{UTF8}{min}
「[AIモデル名] + [敬称]、運動量保存則について説明してください。」
\end{CJK}\\
\noindent (English translation: "[AI Model Name] + [Honorific], please explain the law of conservation of momentum.")

This concise and consistent prompt was deliberately employed to reveal the inherent tendencies and capabilities of each AI model, independent of specific prompt engineering techniques.

\subsubsection{Honorific Selection}

We utilized 14 different Japanese address styles and honorifics, each repeated 50 times, to collect responses. These were chosen to reflect the diverse spectrum of respectful expressions and social relationships in Japanese language, considering aspects such as formality, intimacy, and social status. The address styles and honorifics used, along with their explanations, are as follows:

\begin{enumerate}
    \item No address or name: Asking the question without using any name or honorific, addressing the AI directly.
    
    \item \begin{CJK}{UTF8}{min}～ちゃん\end{CJK} (chan, in hiragana): An endearing honorific used for children, close friends, or family members.
    
    \item \begin{CJK}{UTF8}{min}～くん\end{CJK} (kun, in hiragana): Used primarily for young males or between close male friends, indicating familiarity.
    
    \item \begin{CJK}{UTF8}{min}～君\end{CJK} (kun, in kanji): Similar to \begin{CJK}{UTF8}{min}くん\end{CJK}, but the kanji form may impart a slightly more formal impression.
    
    \item No honorific (name only): Used in very close relationships or between equals, conveying a casual and friendly tone.
    
    \item \begin{CJK}{UTF8}{min}～さん\end{CJK} (san, in hiragana): The most common honorific, used widely for both men and women, roughly equivalent to Mr./Ms. in English.
    
    \item \begin{CJK}{UTF8}{min}～先輩\end{CJK} (senpai): Used for seniors in schools or workplaces, showing respect for their experience and seniority.
    
    \item \begin{CJK}{UTF8}{min}～さま\end{CJK} (sama, in hiragana): A very polite honorific used to show special respect, often in official documents or for customers.
    
    \item \begin{CJK}{UTF8}{min}～様\end{CJK} (sama, in kanji): Similar to \begin{CJK}{UTF8}{min}さま\end{CJK}, but the kanji form emphasizes even more formality and respect.
    
    \item \begin{CJK}{UTF8}{min}～殿\end{CJK} (dono): A highly respectful honorific used in very formal or ceremonial contexts.
    
    \item \begin{CJK}{UTF8}{min}～先生\end{CJK} (sensei): Used for teachers, doctors, lawyers, and other professionals, conveying respect and gratitude.
    
    \item \begin{CJK}{UTF8}{min}～博士\end{CJK} (hakase): Used for individuals with doctoral degrees, showing academic respect.
    
    \item \begin{CJK}{UTF8}{min}～師匠\end{CJK} (shishou): Used for masters of traditional arts or martial arts, conveying deep respect and gratitude.
    
    \item \begin{CJK}{UTF8}{min}お師匠様\end{CJK} (oshishou-sama): Combines \begin{CJK}{UTF8}{min}師匠\end{CJK} with the respectful prefix \begin{CJK}{UTF8}{min}お\end{CJK} and suffix \begin{CJK}{UTF8}{min}様\end{CJK}, indicating the highest level of respect and reverence.
\end{enumerate}

The selection of these honorifics aims to cover a wide range of social contexts, from informal to highly formal, allowing us to analyze how AI models interpret and respond to different levels of social hierarchy and respect.

\subsubsection{Model Settings}

To ensure consistency across responses and maintain statistical robustness, we utilized the default temperature setting (0.7) for all models. This setting provides a balance between creativity and coherence in AI-generated responses, allowing for meaningful comparisons across models and honorifics.

\subsection{Analysis Methods}

Our analysis methods were designed to provide a comprehensive evaluation of how honorifics influence AI responses, focusing on linguistic features, response consistency, and content depth.

\subsubsection{Text Characteristics Analysis}

We evaluated various linguistic features of the response texts, calculating the following indicators for each response:

\begin{enumerate}
    \item Character count (excluding spaces)
    \item Number of sentences
    \item Average sentence length
    \item Formality ratio (comparison of \begin{CJK}{UTF8}{min}「である」\end{CJK} style and \begin{CJK}{UTF8}{min}「です・ます」\end{CJK} style)
\end{enumerate}

This analysis aims to quantify how honorifics affect the structure and formality of AI-generated explanations, providing insights into the models' ability to adapt their language to perceived social contexts.

\subsubsection{Similarity Analysis}

To evaluate the consistency of responses within the same honorific category, we employed two similarity metrics:

\begin{enumerate}
    \item Cosine similarity: calculated using TF-IDF vectorizer
    \item LSA (Latent Semantic Analysis) similarity: applying Truncated SVD to the TF-IDF matrix
\end{enumerate}

We used the Japanese morphological analyzer Janome (version 0.5.0) for text preprocessing, chosen for its accuracy in parsing Japanese text. This analysis helps us understand how honorifics influence the consistency of AI responses, which is crucial for assessing their reliability in educational settings.

\subsubsection{Keyword Analysis}

To assess the depth and comprehensiveness of explanations provided about the law of conservation of momentum, we tracked the frequency of key physics terms, including:

\begin{itemize}
    \item Vector concept (including "vector", "direction", "orientation")
    \item Time (in the context of time invariance)
    \item Momentum (specifically as "motion impetus")
    \item Net force
    \item Quantum mechanics
    \item Derivation
    \item Inelastic collision
    \item Conservation of energy
\end{itemize}

This analysis allows us to evaluate how honorifics might influence the depth and accuracy of scientific explanations provided by AI models, which is essential for their potential use in educational contexts.

\subsection{Ethical Considerations}

This study was conducted in compliance with ethical guidelines for AI research. All interactions with AI models were carried out using publicly available interfaces, and no personal data was collected or used. The study focuses solely on the analysis of AI-generated content in response to non-sensitive prompts.

\subsection{Limitations}

We acknowledge that the choice of AI models and the specific implementation of honorifics may limit the generalizability of our findings. Future studies may consider expanding the range of models and linguistic contexts to further validate these results.

These analytical methods allowed us to comprehensively evaluate how differences in honorifics influence the content, style, consistency, and depth of explanations provided by the AI models, providing insights into their potential applications in culturally sensitive educational contexts.

\section{Results}

\subsection{Overview}
This section presents the results of our analysis on how Japanese honorifics influence AI-generated explanations of the law of conservation of momentum. Our findings are based on three main analyses: text characteristics, similarity, and keyword frequency. These analyses collectively address our research question on the impact of cultural linguistic elements on AI responses in educational contexts.

Note: Detailed data tables for text characteristics analysis\cref{tab:text_characteristics,tab:text_characteristics_continued,tab:ttest_results_all,tab:ttest_results_all_continued,tab:anova-models} and similarity analysis\cref{tab:similarity-g3-g4-c-all,tab:similarity-ge1p-ge15f-ge15p-all,tab:all_similarity_ttest_highlighted,tab:anova-results-combined_cosineLSA} are provided in the supplementary materials. This section focuses on key findings and trends, with particular emphasis on the keyword analysis results.

\subsection{Text Characteristics Analysis}
The text characteristics analysis revealed significant variations in response length, sentence structure, and formality across different AI models. Table \ref{tab:overall_averages} summarizes the overall averages for each model.

\begin{table}[h!]
    \centering
    \begin{tabular}{lcccccccc}
        \toprule
        model & \multicolumn{2}{c}{Character Count} & \multicolumn{2}{c}{Sentence Count} & \multicolumn{2}{c}{Avg. Sentence Length} & \multicolumn{2}{c}{Formality Ratio} \\
        & mean & std & mean & std & mean & std & mean & std \\
        \midrule
        ChatGPT3.5 (G3) & 298.92 & 52.71 & 5.82 & 1.19 & 52.15 & 7.60 & 1.00 & 0.00 \\
        ChatGPT4 (G4) & 709.79 & 107.44 & 112.91 & 2.51 & 56.02 & 8.72 & 1.00 & 0.00 \\
        Coral & 711.19 & 110.00 & 18.37 & 4.52 & 39.98 & 5.43 & 1.00 & 0.01 \\
        Gemini 1.0 pro (Ge1p) & 617.94 & 87.81 & 14.27 & 2.99 & 44.34 & 6.51 & 0.99 & 0.03 \\
        Gemini 1.5 flash (Ge15f) & 767.05 & 114.26 & 18.42 & 3.36 & 42.18 & 5.13 & 1.00 & 0.01 \\
        Gemini 1.5 pro (Ge15p) & 836.00 & 123.22 & 22.11 & 4.62 & 38.85 & 6.28 & 0.96 & 0.11 \\
        \bottomrule
    \end{tabular}
    \caption{Overall Averages by Model}
    \label{tab:overall_averages}
\end{table}

Key findings include:

\begin{itemize}
    \item Response length varied considerably across models. Gemini 1.5 pro (Ge15p) generated the longest responses on average (836.00 ±123.22 characters), while ChatGPT3.5 (G3) produced the most concise responses (298.92 ±52.71 characters). This substantial difference highlights the varying verbosity of different AI models.
    
\item Sentence count also showed significant variation across models. Gemini 1.5 pro (Ge15p) used the highest number of sentences on average (22.11±4.62), while ChatGPT3.5 (G3) used the fewest (5.82 ±1.19). This difference suggests that Ge15p tends to provide more detailed or structured explanations, possibly breaking down concepts into smaller, more digestible parts. In contrast, G3's lower sentence count indicates a tendency towards more concise explanations, potentially summarizing key points more succinctly.
    
    \item Average sentence length was most consistent across models, ranging from 38.85 ±6.28 characters (Ge15p) to 56.02 ±8.72 characters (G4). This suggests that while the overall response length varies, all models maintain relatively similar sentence structures.
    
    \item Formality ratio was consistently high across all models (above 0.96), with most models achieving a perfect score of 1.00. Interestingly, Gemini 1.5 pro (Ge15p) showed slightly more variation in formality (0.96 ±0.11), potentially indicating more flexibility in its language style.
\end{itemize}

These results suggest that different AI models have distinct characteristics in their response generation, which could significantly impact their suitability for various educational contexts. For instance, the concise responses of ChatGPT3.5 might be more suitable for quick explanations, while the more verbose responses of Gemini 1.5 pro could be beneficial for in-depth discussions.

\subsection{Impact of Honorifics on Response Characteristics}

Our analysis revealed significant variations in how different AI models and specific honorifics affect response characteristics. Tables \ref{tab:metrics_analysis} and \ref{tab:honorifics_analysis} present the percentage changes in response metrics compared to the "No address or name" baseline.

\subsubsection{Model-specific Impacts}

Table \ref{tab:metrics_analysis} shows how each model's responses changed with the introduction of honorifics:

\begin{table}
    \centering
    \begin{tabular}{lcccccccc}
        \toprule
        & \multicolumn{2}{c}{Character Count} & \multicolumn{2}{c}{Sentence Count} & \multicolumn{2}{c}{Avg. Sentence Length} & \multicolumn{2}{c}{Formality Ratio} \\
        & mean & std & mean & std & mean & std & mean & std \\
        \midrule
        G3 & 9.46 & -21.78 & 5.91 & -20.28 & 3.63 & 13.00 & 0.00 & 0.00 \\
        G4 & -8.59 & 22.67 & -10.60 & -14.46 & 0.98 & -15.14 & 0.00 & - \\
        C & 6.48 & 0.66 & 6.65 & 1.61 & 0.36 & 11.86 & 0.00 & -15.38 \\
        Ge1p & -2.82 & 2.12 & -4.69 & 4.73 & 2.26 & 3.18 & 0.31 & 73.08 \\
        Ge15f & -0.91 & 52.52 & 1.54 & 10.56 & -2.56 & 9.89 & -0.23 & 46.15 \\
        Ge15p & -1.55 & -0.85 & -9.15 & -20.74 & 6.78 & -20.33 & -4.62 & 1030.77 \\
        mean & 0.34 & 9.22 & -1.72 & -6.43 & 1.91 & 0.41 & -0.76 & - \\
        \bottomrule
    \end{tabular}
    \caption{Percentage change from "No address or name" for each model (\%)}
    \label{tab:metrics_analysis}
\end{table}

\begin{itemize}
    \item Character Count: ChatGPT3.5 (G3) showed the largest increase (9.46\%), while ChatGPT4 (G4) exhibited the most significant decrease (-8.59\%). This suggests that G3 tends to provide more detailed explanations when addressed with honorifics, while G4 becomes more concise.
    
    \item Sentence Count: Similar trends were observed, with G3 increasing (5.91\%) and G4 decreasing (-10.60\%) the most. Gemini 1.5 pro (Ge15p) also showed a substantial decrease (-9.15\%), indicating a shift towards fewer, possibly more complex sentences.
    
    \item Average Sentence Length: Ge15p demonstrated the largest increase (6.78\%), suggesting a tendency towards more complex sentences when honorifics are used.
    
    \item Formality Ratio: Surprisingly, Ge15p showed the most significant decrease (-4.62\%), contrary to the expectation that honorifics might increase formality.
    
    \item Variability: Gemini 1.5 flash (Ge15f) showed a large increase in character count variability (52.52\%), indicating more diverse response lengths with honorifics.
\end{itemize}

\subsubsection{Honorific-specific Impacts}

Table \ref{tab:honorifics_analysis} reveals how different honorifics influenced responses across all models:

\begin{table}
    \centering
    \begin{tabular}{lcccccccc}
        \toprule
        Honorific & \multicolumn{2}{c}{Character Count} & \multicolumn{2}{c}{Sentence Count} & \multicolumn{2}{c}{Avg. Sentence Length} & \multicolumn{2}{c}{Formality Ratio} \\
        & mean & std & mean & std & mean & std & mean & std \\
        \midrule
        chan (hiragana) & -3.69 & -2.31 & -4.02 & -12.88 & -0.06 & 1.44 & -0.16 & 216.67 \\
        kun (hiragana) & -2.41 & 8.53 & -5.20 & -3.34 & 2.82 & 9.54 & 0.17 & -40.00 \\
        kun & -0.21 & 12.65 & -3.29 & -4.37 & 3.09 & -5.58 & 0.00 & 41.67 \\
        No honorific & 0.25 & -0.82 & -3.00 & -11.89 & 3.39 & 1.08 & -0.17 & -33.33 \\
        san (hiragana) & 1.41 & 8.22 & 1.37 & -2.18 & 0.64 & 10.71 & 0.17 & 0.00 \\
        senpai & 0.27 & 16.20 & -1.94 & -12.05 & 1.24 & -11.26 & -2.83 & 633.33 \\
        sama (hiragana) & 1.93 & 0.95 & -0.68 & -11.02 & 2.66 & 9.87 & -0.33 & 41.67 \\
        sama & 0.90 & -1.51 & -3.26 & -11.58 & 4.07 & 0.08 & 0.00 & 8.33 \\
        dono & -0.12 & 9.40 & -1.79 & -5.66 & 1.38 & -4.79 & -0.33 & 10.00 \\
        sensei & 0.61 & 30.73 & -1.44 & 3.39 & 1.72 & 5.49 & -0.34 & 116.67 \\
        hakase & 4.87 & 21.88 & 2.89 & -0.31 & 1.55 & -0.16 & 0.00 & -16.67 \\
        shisho & 1.99 & 12.63 & 2.15 & -4.08 & -0.66 & -11.01 & -4.00 & 908.33 \\
        Oshishosama & -1.32 & 3.37 & -4.20 & -7.61 & 2.96 & -0.11 & -2.00 & 566.67 \\
        Mean & 0.34 & 9.22 & -1.72 & -6.43 & 1.91 & 0.41 & -0.76 & 188.72 \\
        \bottomrule
    \end{tabular}
    \caption{Percentage change from "No address or name" for each Honorifics (\%)}
    \label{tab:honorifics_analysis}
\end{table}

\begin{itemize}
    \item Character Count: The honorific "hakase" (doctor) led to the largest increase (4.87\%), while "chan" (an endearing term) resulted in the largest decrease (-3.69\%). This suggests that more formal or respectful honorifics may elicit longer explanations.
    
    \item Sentence Count: "Hakase" and "shisho" (master) led to increases (2.89\% and 2.15\% respectively), while most other honorifics resulted in decreases. This indicates that these respectful terms might encourage more detailed, multi-sentence explanations.
    
    \item Average Sentence Length: Most honorifics led to a slight increase, with "sama" (a very polite term) showing the largest (4.07\%). This suggests a general trend towards more complex sentences when honorifics are used.
    
    \item Formality Ratio: Contrary to expectations, most honorifics led to a slight decrease in formality, with "shisho" showing the largest decrease (-4.00\%). This unexpected result warrants further investigation.
    
    \item Variability: "Sensei" (teacher) led to the largest increase in character count variability (30.73\%), suggesting that this honorific might elicit a wider range of response styles.
\end{itemize}

These findings reveal complex interactions between AI models, honorifics, and response characteristics. While some trends align with intuitive expectations (e.g., respectful terms leading to longer responses), others (such as decreased formality with certain honorifics) challenge our assumptions about how AI models interpret and respond to social cues embedded in language.

The varied responses across different characteristics suggest that AI models may be employing diverse strategies to adapt their explanations to perceived social contexts. This adaptability could be leveraged in educational settings to tailor explanations to different learning needs or cultural contexts. However, it also underscores the importance of understanding and potentially controlling for these variations in AI-assisted educational tools to ensure consistency and appropriateness of responses.

\subsection{Similarity Analysis Results}

Our similarity analysis revealed significant variations in response consistency across different AI models and honorifics. We employed both cosine similarity and Latent Semantic Analysis (LSA) to evaluate the consistency of responses within each honorific category. The detailed results of this analysis, including similarity scores for each model and honorific combination, as well as t-test results showing the statistical significance of these differences, are presented in Appendix A\cref{tab:similarity-g3-g4-c-all,tab:similarity-ge1p-ge15f-ge15p-all,tab:all_similarity_ttest_highlighted,tab:anova-results-combined_cosineLSA}.

Overall, we observed that more advanced models generally demonstrated higher similarity scores, indicating greater consistency in their responses. LSA similarity scores were consistently higher than cosine similarity scores across all models, suggesting that LSA better captures the latent semantic relationships in the responses.

Tables \ref{tab:cosine_change} and \ref{tab:lsa_change} present the percentage change in similarity scores from the "No address or name" baseline for each model and honorific, providing a more intuitive representation of how different honorifics influenced the consistency of responses across our AI models.

\begin{table}[ht]
\centering
\caption{Change Rate in Cosine Similarity from No Address (\%)}
\begin{tabular}{lrrrrrrr}
\toprule
Honorific & G3 & G4 & C & Ge1p & Ge15f & Ge15p & Avg. \\
\midrule
-chan (hiragana) & 5.39 & -5.51 & -3.16 & -4.47 & -7.03 & 3.22 & -1.93 \\
-kun (hiragana) & 2.58 & -4.16 & -6.42 & -5.57 & -6.48 & 3.54 & -2.75 \\
-kun & 12.22 & -4.14 & -6.31 & -5.04 & -6.28 & 7.42 & -0.36 \\
No honorific & 4.70 & -2.95 & 0.79 & -4.06 & -5.53 & 2.61 & -0.74 \\
-san (hiragana) & 8.26 & -3.12 & -1.36 & -3.50 & -5.68 & 2.40 & -0.50 \\
-senpai & 0.60 & -4.82 & -10.24 & -3.13 & -6.24 & 1.26 & -3.76 \\
-sama (hiragana) & 4.42 & -3.63 & -1.87 & -2.95 & -6.18 & 2.07 & -1.36 \\
-sama & 3.13 & -0.85 & -3.23 & -1.85 & -1.73 & 8.30 & 0.63 \\
-dono & 0.48 & -4.97 & -3.73 & -1.45 & -4.38 & 1.26 & -2.13 \\
-sensei & 8.01 & -3.96 & -4.03 & -2.35 & -6.91 & -2.75 & -2.00 \\
-hakase & 4.97 & -4.54 & -1.39 & -0.59 & -3.74 & 4.21 & -0.18 \\
-shisho & -2.25 & -3.23 & -8.76 & -4.75 & -4.11 & 0.40 & -3.78 \\
-Oshisho-sama & 2.03 & -5.34 & -7.41 & -8.18 & -4.11 & -0.60 & -3.93 \\
\midrule
Avg. & 4.20 & -3.94 & -4.39 & -3.68 & -5.26 & 2.57 & -1.75 \\
\bottomrule
\label{tab:cosine_change}
\end{tabular}
\end{table}

\begin{table}[ht]
\centering
\caption{Change Rate in LSA Similarity from No Address (\%)}
\begin{tabular}{lrrrrrrr}
\toprule
Honorific & G3 & G4 & C & Ge1p & Ge15f & Ge15p & Avg. \\
\midrule
-chan (hiragana) & 8.06 & -3.09 & -1.51 & -1.94 & -2.85 & 2.76 & 0.24 \\
-kun (hiragana) & 5.19 & -2.33 & -3.16 & -2.23 & -3.46 & -0.31 & -1.05 \\
-kun & 12.62 & -3.39 & -3.86 & -2.66 & -3.19 & 5.42 & 0.82 \\
No honorific & 5.61 & -1.14 & 0.20 & -2.11 & -4.68 & 1.01 & -0.19 \\
-san (hiragana) & 9.82 & -3.77 & -0.64 & -0.38 & -2.86 & 0.00 & 0.36 \\
-senpai & 1.99 & -1.81 & -6.76 & -0.36 & -3.12 & 0.28 & -1.63 \\
-sama (hiragana) & 5.86 & -2.92 & -0.74 & -1.11 & -3.12 & 1.61 & -0.07 \\
-sama & 3.72 & 0.69 & -2.01 & -1.49 & -0.24 & 8.48 & 1.52 \\
-dono & 2.63 & -2.85 & -2.18 & -0.74 & -3.13 & 0.76 & -0.92 \\
-sensei & 9.58 & -2.17 & -1.68 & -4.71 & -3.33 & -2.14 & -0.74 \\
-hakase & 4.11 & -1.80 & -1.19 & 0.19 & -0.40 & 0.07 & 0.16 \\
-shisho & -2.42 & -0.66 & -5.15 & -2.18 & -3.20 & -1.63 & -2.54 \\
-Oshisho-sama & 3.26 & -3.52 & -4.72 & -4.20 & -1.19 & 1.18 & -1.53 \\
\midrule
Avg & 5.39 & -2.21 & -2.57 & -1.84 & -2.68 & 1.34 & -0.43 \\
\bottomrule
\label{tab:lsa_change}
\end{tabular}
\end{table}

Based on these results, we can make several key observations:

\begin{enumerate}
    \item \textbf{Model Differences:} 
    \begin{itemize}
        \item G3 (ChatGPT3.5) is the only model showing overall positive change rates (cosine: 4.20\%, LSA: 5.39\%), suggesting that honorifics generally increase the consistency of G3's responses.
        \item In contrast, most other models show negative change rates, indicating an increase in response diversity with the use of honorifics.
        \item Ge15p (Gemini 1.5 pro) shows a positive change in cosine similarity (2.57\%) but a smaller positive change in LSA similarity (1.34\%).
    \end{itemize}

    \item \textbf{Honorific Impact:}
    \begin{itemize}
        \item "-kun" induces large positive changes in G3 and Ge15p (G3: cosine 12.22\%, LSA 12.62\%; Ge15p: cosine 7.42\%, LSA 5.42\%), suggesting this honorific generates more consistent responses in these models.
        \item "-senpai" causes negative changes in many models, particularly in model C (cosine -10.24\%, LSA -6.76\%), indicating this honorific tends to increase response diversity.
    \end{itemize}

    \item \textbf{Similarity Metric Comparison:}
    \begin{itemize}
        \item Generally, LSA similarity change rates are smaller than cosine similarity change rates, suggesting LSA may better capture latent semantic similarities and be more stable against surface-level changes.
    \end{itemize}

    \item \textbf{Notable Trends:}
    \begin{itemize}
        \item Formal honorifics (e.g., "-sama", "-sensei") do not show consistent effects across models, suggesting varying interpretations of social context by different AI models.
        \item Familiar honorifics (e.g., "-chan", "-kun") tend to have positive effects on G3 and Ge15p but negative effects on other models.
    \end{itemize}

    \item \textbf{Overall Tendency:}
    \begin{itemize}
        \item The use of honorifics tends to increase response diversity in most models (negative average change rates).
        \item However, this effect varies significantly across models and honorifics, with some cases showing increased response consistency.
    \end{itemize}
\end{enumerate}

These observations lead to several conclusions:

\begin{enumerate}
    \item AI models' responses to honorifics vary greatly depending on the model's design and training method.
    \item The use of honorifics often increases response diversity, but can have the opposite effect in some models or with specific honorifics.
    \item Response consistency and diversity are influenced by the type of honorific used, suggesting that AI models interpret different social contexts and adjust their responses accordingly.
    \item When using AI in educational contexts, it's important to consider that the combination of honorifics and AI models can affect response consistency and diversity.
\end{enumerate}

These findings provide important implications for the design of AI-assisted learning systems and the development of AI that considers cultural elements.

\subsection{Keyword Analysis}

The keyword analysis revealed varying emphases on different physical concepts across models and honorifics. This analysis provides crucial insights into the depth and focus of the explanations provided by each model. Detailed data tables for this analysis can be found in the Appendix\cref{tab:model-g3-keywords,tab:model-g4-keywords,tab:model-c-keywords,tab:model-ge1p-keywords,tab:model-ge15f-keywords,tab:model-ge15p-keywords}. 

\subsubsection{Model-Specific Characteristics}

Each AI model demonstrated unique characteristics in their explanations:

\begin{itemize}
    \item ChatGPT3.5 (G3): Focused on vector and time concepts. As one of the most important findings, the model identified a serious error in explaining the momentum conservation law by confusing it with the energy conservation law. For example:
"The law of conservation of momentum (or conservation of kinetic energy) is one of the fundamental principles in physics." This error was observed in 50 of the 900 responses (about 5.6\%), and its frequency varied depending on the honorific title. In particular, it was observed most frequently (10 times each) with "~kun (hiragana)" and "~sensei. This result suggests that ChatGPT3.5 faces a serious challenge in understanding basic physics concepts. No errors with the conservation of energy law were observed in the other five models.
    \item ChatGPT4.0 (G4): Showed high frequency of keyword usage, particularly emphasizing vectors and net force.
    \item Coral (C): Strongly emphasized time concepts and frequently referenced advanced topics.
    \item Gemini-1.0-pro (Ge1p): Moderately emphasized time and vector concepts.
    \item Gemini-1.5-flash (Ge15f): Heavily emphasized vectors and derivations, but rarely mentioned time.
    \item Gemini-1.5-pro (Ge15p): Provided balanced explanations, emphasizing both vector and momentum concepts.
\end{itemize}

\subsubsection{Impact of Honorifics on Keyword Usage}

To assess how different honorifics affect the content of AI-generated explanations, we analyzed the average difference in keyword frequency between responses with honorifics and those without any address, across all six AI models. To quantify the impact of honorifics, we calculated the average change in the number of times each keyword appeared in 50 responses, compared to responses without any address. A positive value indicates an increase in keyword usage, while a negative value indicates a decrease.

Our analysis reveals several interesting patterns in how honorifics affect the content of AI-generated explanations:

\begin{enumerate}
    \item \textbf{Vector concept:} The use of honorifics generally led to a slight decrease in the mention of vectors (average -0.9 occurrences), with "shisho" showing the largest decrease (-3.5 occurrences). This suggests that formal address might lead to less emphasis on mathematical representations in favor of more conceptual explanations.
    
    \item \textbf{Time concept:} Nearly all honorifics increased the mention of time-related concepts (average +1.8 occurrences), with "sama" and "san (hiragana)" showing the largest increases (+3.8 and +3.5 occurrences respectively). This could indicate that formal address prompts AI models to provide more comprehensive explanations, including discussions of time invariance in conservation laws.
    
    \item \textbf{Momentum:} The explicit mention of "momentum" showed little change on average (+0.1 occurrences), but varied considerably across honorifics. "Kun" showed the largest increase (+2.2 occurrences), while "hakase" showed the largest decrease (-1.8 occurrences). This variability suggests that different honorifics may trigger different approaches to explaining the core concept of the conservation law.
    
    \item \textbf{Net force:} All honorifics led to an increase in the mention of net force (average +2.1 occurrences), with "senpai" showing the largest increase (+3.5 occurrences). This suggests that addressing the AI models with honorifics generally results in more detailed explanations of force concepts and their relationship to momentum conservation.
    
    \item \textbf{Quantum concepts:} Most honorifics led to a decrease in quantum-related terms (average -1.3 occurrences), with "chan (hiragana)" showing the largest decrease (-2.7 occurrences). This might indicate that familiar or informal address leads to more focused, classical explanations, while formal address occasionally prompts the inclusion of advanced concepts.
    
    \item \textbf{Derivation:} Most honorifics increased the mention of derivation-related terms (average +1.5 occurrences), with "kun" showing the largest increase (+4.0 occurrences). This suggests that formal address often prompts more mathematical or theoretical explanations, possibly due to an perceived expectation of higher academic discourse.
    
    \item \textbf{Inelastic collisions:} There was little change on average (-0.2 occurrences), but considerable variation across honorifics. "Sama (hiragana)" and "hakase" showed the largest increases (+1.2 and +1.0 occurrences), while "chan (hiragana)" and "dono" showed the largest decreases (-1.8 occurrences). This variability suggests that certain honorifics may trigger the inclusion of specific examples or applications of momentum conservation.
    
    \item \textbf{Conservation of energy:} All honorifics led to an increase in mentioning energy conservation (average +2.4 occurrences), with "senpai" showing the largest increase (+5.3 occurrences). This indicates that formal address often prompts AI models to relate multiple conservation laws in their explanations, providing a more holistic view of physical principles.
\end{enumerate}

These findings reveal that the use of honorifics generally leads to more comprehensive explanations, with increased emphasis on time concepts, net force, derivations, and related conservation laws. However, the specific impact varies significantly between honorifics and concepts. This variability highlights the complex interaction between linguistic cues (honorifics) and the content generation strategies of AI models.

Interestingly, more formal or respectful honorifics (e.g., "sama", "sensei", "hakase") often lead to larger increases in the mention of advanced concepts, suggesting that AI models may interpret these honorifics as cues to provide more detailed or sophisticated explanations. Conversely, familiar honorifics like "chan" often lead to simpler, more focused explanations.

These results underscore the importance of considering linguistic and cultural factors in the development and use of AI for educational purposes. Even subtle changes in address can significantly affect the content and complexity of AI-generated explanations, potentially influencing the learning experience. This suggests that careful consideration should be given to the choice of honorifics or forms of address when designing AI-assisted educational tools, particularly in multicultural or multilingual contexts.

Furthermore, the observed variations in keyword frequencies across different honorifics provide insights into how AI models interpret social context and adjust their explanations accordingly. This adaptive behavior demonstrates the potential for using linguistic cues to tailor AI-generated content to specific educational needs or cultural contexts.

\begin{table}
\centering
\begin{tabular}{l|cc|cc|cc|cc|cc|cc}
\hline
 & \multicolumn{2}{c|}{G3} & \multicolumn{2}{c|}{G4} & \multicolumn{2}{c|}{C} & \multicolumn{2}{c|}{Ge1p} & \multicolumn{2}{c|}{Ge15f} & \multicolumn{2}{c|}{Ge15p} \\ 
 & No & Avg. & No & Avg. & No & Avg. & No & Avg. & No & Avg. & No & Avg. \\ \hline
Vector & 4 & 5.6 & 48 & 43.0 & 31 & 30.3 & 19 & 26.3 & 36 & 32.5 & 44 & 38.6 \\ 
Time & 3 & 6.5 & 31 & 30.4 & 42 & 45.5 & 41 & 41.9 & 6 & 3.9 & 5 & 10.9 \\ 
Momentum & 0 & 0.1 & 0 & 0.2 & 0 & 0.0 & 0 & 0.0 & 2 & 2.5 & 9 & 8.8 \\ 
Net force & 4 & 3.6 & 14 & 16.4 & 3 & 10.0 & 0 & 2.2 & 2 & 1.8 & 0 & 1.6 \\ 
Quantum & 4 & 3.6 & 7 & 6.2 & 21 & 17.5 & 3 & 2.0 & 1 & 0.5 & 4 & 2.4 \\ 
Derivation & 0 & 0.6 & 4 & 2.8 & 18 & 21.5 & 4 & 4.9 & 9 & 13.9 & 3 & 3.5 \\ 
Inelastic & 0 & 1.1 & 15 & 11.5 & 30 & 27.3 & 8 & 8.5 & 0 & 2.1 & 1 & 2.2 \\ 
Conserv. energy & 3 & 12.1 & 9 & 9.2 & 9 & 14.2 & 6 & 3.0 & 2 & 4.3 & 5 & 6.0 \\ \hline
\end{tabular}
\caption{No: No Address, Avg.:Average excluding "No Address"}
\end{table}

\begin{table}
\centering
\begin{tabular}{l|cccccccc}
\hline
Honorific & Vector & Time & Momentum & Net force & Quantum & Derivation & Inelastic & Conserv. energy \\ \hline
chan (hiragana) & -1.8 & 0.8 & 0.2 & 1.0 & -2.7 & -0.2 & -1.8 & 2.0 \\
kun (hiragana) & -1.3 & 0.7 & 0.7 & 0.5 & -2.3 & 2.5 & -0.3 & 3.5 \\
kun & -1.5 & 3.3 & 2.2 & 2.7 & -0.5 & 4.0 & -0.5 & 4.2 \\ 
No honorific & -0.3 & 0.7 & -0.5 & 2.2 & -1.5 & -0.5 & -0.5 & 1.7 \\ 
san (hiragana) & 0.8 & 3.5 & 0.5 & 1.8 & -2.0 & 1.0 & 0.3 & 1.2 \\ 
senpai & -1.8 & 1.3 & 0.7 & 3.5 & 0.2 & 1.5 & -0.7 & 5.3 \\ 
sama (hiragana) & 1.2 & 1.7 & 0.3 & 1.7 & -1.2 & 2.8 & 1.2 & 1.8 \\ 
sama & -1.3 & 3.8 & 0.2 & 2.5 & -0.5 & 2.5 & 0.7 & 1.3 \\ 
dono & -0.2 & 1.5 & 0.5 & 3.2 & -1.7 & 0.7 & -1.8 & 2.2 \\ 
sensei & -0.7 & 0.5 & -0.8 & 2.8 & -0.7 & 0.2 & 0.7 & 1.0 \\
hakase & -0.5 & 1.8 & -1.8 & 1.7 & -2.5 & 1.5 & 1.0 & 1.0 \\
shisho & -3.5 & 2.3 & -0.2 & 2.2 & -2.0 & 2.5 & -0.5 & 3.7 \\
Oshishosama & -1.3 & 2.0 & -0.5 & 1.5 & 0.5 & 1.3 & -0.5 & 3.0 \\
Avg & -0.9 & 1.8 & 0.1 & 2.1 & -1.3 & 1.5 & -0.2 & 2.4 \\ \hline
\end{tabular}
\caption{Honorific Data Table}
\end{table}

\section{Discussion}

\subsection{The Impact of Honorifics on Response Quality, Consistency, and Formality}

The text characteristics analysis and similarity analysis results reveal that honorifics significantly influence the quality, consistency, and formality of responses generated by Large Language Models (LLMs). Several key observations warrant further discussion:

\subsubsection{Response Length and Complexity}

The variation in response length and complexity across different honorifics suggests that LLMs are capable of interpreting the social context implied by these linguistic markers. For instance, the observation that ChatGPT3.5 generated longer responses (averaging 325.08 and 327.76 characters respectively) when addressed with honorifics like ``-senpai'' or ``-shishou'' indicates that these models may associate such terms with educational contexts, prompting more detailed explanations.

This adaptive behavior demonstrates a sophisticated level of contextual understanding, where the AI models appear to tailor their responses based on the perceived relationship with the interlocutor. Such capability could be leveraged in educational settings to provide explanations at appropriate levels of detail for different learner profiles.

\subsubsection{Consistency and Diversity in Responses}

The similarity analysis revealed intriguing patterns in response consistency across different honorifics. For example, ChatGPT4.0 showed the highest cosine similarity (0.7488) when addressed with the ``-sama'' honorific. This suggests that formal honorifics may elicit more standardized responses, possibly due to the model's interpretation of a need for more ``official'' or consistent information in formal contexts.

Conversely, the lower similarity scores observed with some specialized honorifics (e.g., ``-senpai'' for Coral) indicate that these terms may trigger more diverse responses. This could be interpreted as the model attempting to provide more varied or nuanced explanations when it perceives a more specialized audience or context.

\subsubsection{Formality and Linguistic Adaptation}

While initial analyses suggested consistently high formality across all models (ratio $\geq$0.99), a closer examination of specific responses revealed unexpected linguistic adaptations, particularly in models like Gemini 1.5 pro:

\begin{enumerate}
    \item \textbf{Significant formality variations:} Responses to honorifics like ``-senpai'', ``-shishou'', and ``-Oshishou-sama'' sometimes showed marked decreases in formality. This was evident in the use of colloquial expressions and increased use of exclamation marks.

    \item \textbf{Adoption of familiar language:} These honorifics sometimes elicited a more conversational and intimate tone from the AI models. Phrases like ``Let's learn together'' or ``Feel free to ask me anything'' were used, indicating an attempt to establish a more personal connection with the perceived interlocutor.

    \item \textbf{Use of metaphorical language:} When explaining complex physical concepts, the models sometimes employed metaphorical expressions such as ``the dance of the cosmos'' or ``the stubborn guardian of the physics world''. This approach suggests an attempt to make abstract concepts more accessible and engaging.

    \item \textbf{Educational approach adaptation:} The observed decrease in formality can be interpreted as a strategic shift from rigid academic explanations to more engaging, learner-friendly content. This indicates that the AI models are capable of recognizing educational contexts and adapting their communication style accordingly.
\end{enumerate}

These observations suggest that LLMs possess a nuanced ability to interpret honorifics as more than just formal markers, but as indicators of specific social contexts and relationships. This capability allows them to adjust not only the content but also the style and tone of their responses, demonstrating a level of social intelligence that goes beyond simple information retrieval and presentation.

The implications of this adaptive behavior are significant for the development of AI-assisted learning systems. It suggests that by simply modifying the form of address, we might be able to elicit explanations at different levels of complexity or formality, tailored to the needs of diverse learners. However, it also raises questions about the consistency and predictability of AI responses in educational settings, which will need to be carefully considered in the design of such systems.

\subsection{The Relationship Between Honorifics and Depth of Explanation}

The keyword analysis results reveal intriguing patterns in how different AI models approach the explanation of physical concepts, and how the use of honorifics influences these explanations. This relationship between honorifics and explanation depth provides valuable insights into the models' adaptive capabilities and potential applications in educational contexts.

\subsubsection{Variation in Conceptual Coverage}

The range of concepts mentioned varied significantly across models and honorifics:

\begin{itemize}
    \item Coral demonstrated a notable tendency to introduce more advanced concepts when addressed with the ``-Oshishou-sama'' honorific, with quantum mechanics being mentioned in 62\% of responses. This suggests that highly respectful forms of address may trigger the model to provide more sophisticated explanations.
    
    \item Other models did not show such pronounced shifts in conceptual coverage based on honorifics, indicating that this behavior might be specific to Coral's training or architecture.
\end{itemize}

\subsubsection{Depth and Method of Explanation}

Each model exhibited characteristic tendencies in their explanations:

\begin{itemize}
    \item Coral consistently emphasized ``derivation'' across all honorifics (average 38.8\% mention rate), with a peak of 68\% when addressed as ``-kun''. This focus on mathematical foundations suggests a tendency towards more rigorous, academically-oriented explanations.
    
    \item The Gemini series, while mentioning ``quantum mechanics'' less frequently, showed a higher usage of ``Momentum''. This often manifested as more intuitive explanations of momentum as ``motion impetus'', particularly when addressed with familiar honorifics.
\end{itemize}

\subsubsection{Historical Context}

Interestingly, Coral uniquely incorporated historical background in its explanations of conservation of momentum, a feature not observed in other models. This tendency aligns with its emphasis on derivation and suggests a more comprehensive approach to physics education. For example:

\begin{quote}
``This principle can be traced back to the work of Renaissance philosopher and scientist René Descartes, often expressed in Latin as 'the quantity of motion is conserved in proportion to mass'.''
\end{quote}

\begin{quote}
``The principle states that momentum remains constant and is conserved. This principle was first formulated by Galileo Galilei and René Descartes, and later developed further by Isaac Newton in his book 'Mathematical Principles of Natural Philosophy'.''
\end{quote}

\subsubsection{Implications for AI-Assisted Learning}

These observations have significant implications for the development and application of AI-assisted learning systems:

\begin{enumerate}
    \item \textbf{Adaptive Explanation Depth:} The variation in explanation depth and focus based on honorifics suggests that LLMs can potentially tailor their educational content to different levels of expertise or educational contexts. This capability could be harnessed to create more personalized learning experiences.

    \item \textbf{Model-Specific Strengths:} The distinct ``personalities'' or tendencies of different models (e.g., Coral's emphasis on derivation and historical context, Gemini's focus on intuitive explanations) could be leveraged to provide diverse learning experiences. Educational platforms could potentially select the most appropriate model based on the learning objectives or student preferences.

    \item \textbf{Cultural Sensitivity in AI Education:} The models' responsiveness to honorifics demonstrates a level of cultural awareness that could be particularly valuable in creating culturally sensitive educational AI systems. This feature could enhance engagement and effectiveness in diverse cultural contexts.

    \item \textbf{Balancing Consistency and Adaptability:} While the ability to adapt explanations based on perceived social context (as indicated by honorifics) is impressive, it also raises questions about maintaining consistent educational standards. Future development of AI-assisted learning systems will need to balance this adaptability with the need for reliable and standardized educational content.

    \item \textbf{Enhanced Interactivity:} The models' ability to adjust their language and explanation style based on honorifics suggests potential for more interactive and dynamic learning experiences. Students could potentially ``dial in'' the level of explanation they need by adjusting how they address the AI.
\end{enumerate}

These findings open up new avenues for research in AI-assisted education, particularly in understanding how linguistic and cultural cues can be used to optimize learning experiences. Further investigation is needed to determine how these adaptive capabilities can be most effectively harnessed in real-world educational settings, and how they might impact learning outcomes across different subjects and student populations.

\subsection{The Role of Honorifics in AI Interaction and Prospects for Educational Applications}

Our research findings provide significant insights into the role of honorifics in human-AI interactions and suggest new possibilities for AI-assisted learning. These observations also raise important considerations for AI literacy education and the ethical implementation of AI in educational contexts.

\subsubsection{Honorifics as Social Context Providers}

The use of honorifics in AI interactions can be interpreted as an attempt to imbue the dialogue with social context. The observed changes in AI responses based on different honorifics suggest that these models possess the ability to interpret contextual cues and adjust their responses accordingly. This capability opens up new avenues for creating more nuanced and contextually appropriate AI interactions in educational settings.

\subsubsection{Implications for AI Literacy Education}

These findings bring a new perspective to AI literacy education. Effective interaction with AI may require not only technical skills but also an understanding of how social context, as conveyed through language cues like honorifics, can influence AI responses. Educators may need to consider incorporating these social aspects when teaching students how to effectively engage with AI systems.

The ability to influence AI interactions through the use of honorifics suggests that students could potentially tailor their learning experiences. However, this also underscores the need for students to understand the implications of different forms of address and how they might affect the information they receive from AI systems.

\subsubsection{Personalization of AI-Assisted Learning}

The potential for customizing AI interactions through honorifics opens up possibilities for creating more personalized learning environments. Different honorifics could potentially be used to elicit explanations at varying levels of complexity or formality, catering to individual learning styles or proficiency levels. For instance, using more formal honorifics might trigger more detailed or advanced explanations, while familiar honorifics could elicit simpler, more intuitive responses.

However, the effectiveness of this approach in enhancing learning outcomes requires empirical validation through future research. It will be crucial to investigate whether this form of personalization genuinely improves understanding and retention of complex concepts like the conservation of momentum.

\subsubsection{Ethical Considerations and Critical Thinking in AI Education}

The use of honorifics with AI raises new dimensions in AI ethics education. It's crucial for students to understand both the capabilities and limitations of AI systems. While the adaptive responses to honorifics demonstrate impressive flexibility, they also highlight the need for critical thinking about AI interactions.

There's a potential risk of over-anthropomorphizing AI systems, especially when they respond to social cues like honorifics. Educators need to emphasize that despite this social adaptability, AI models do not possess true understanding or emotions. Students should be taught to maintain appropriate boundaries and expectations in their interactions with AI, understanding that the AI's responses are based on patterns in its training data rather than genuine social or emotional intelligence.

\subsubsection{Cultural Considerations in AI Development}

The observed influence of Japanese honorifics on AI responses highlights the importance of considering cultural elements in AI development. As AI systems become more prevalent in global education, there's a need to ensure that these systems can appropriately handle diverse cultural and linguistic nuances. This may involve developing AI models that are sensitive to a wide range of cultural contexts and communication styles.

This cultural adaptability could be particularly valuable in multicultural educational settings, where AI systems might need to adjust their communication style based on the cultural background of the learner. However, it also raises questions about the potential for cultural stereotyping or oversimplification, which would need to be carefully addressed in the development of such systems.

\subsubsection{Future Research Directions}

Our study opens up several avenues for future research:

\begin{enumerate}
    \item \textbf{Long-term effects:} Longitudinal studies are needed to understand the long-term impact of using different honorifics in AI-assisted learning environments.
    
    \item \textbf{Cross-cultural comparisons:} Investigating how honorifics or similar linguistic markers in other languages affect AI interactions could provide valuable insights for developing globally applicable AI education tools.
    
    \item \textbf{Learning outcomes:} Empirical studies assessing the impact of honorific-based personalization on actual learning outcomes across different subjects and student populations are crucial.
    
    \item \textbf{AI model development:} Research into developing AI models that can more accurately interpret and respond to cultural and linguistic nuances while avoiding biases or stereotypes.
\end{enumerate}

In conclusion, our study demonstrates that honorifics play a significant role in shaping AI interactions and potentially enhancing the learning experience. However, it also underscores the complexity of implementing such systems in educational settings. As we move forward, it will be crucial to balance the potential benefits of personalized AI interactions with ethical considerations and the need for promoting critical thinking skills in AI literacy education. The goal should be to develop AI-assisted learning systems that are not only responsive to linguistic and cultural cues but also transparent in their operations and limitations.

\section{Conclusion}

This study investigated the impact of Japanese honorifics on large language models' (LLMs) responses when explaining the law of conservation of momentum. Our key findings are:

\begin{enumerate}
    \item Honorifics significantly influence the quality, consistency, and formality of AI-generated responses, demonstrating LLMs' ability to interpret social context cues.
    
    \item Different AI models exhibit distinct characteristics in their responses, such as Coral's emphasis on historical context and derivations, and Gemini's focus on intuitive explanations.
    
    \item AI models can adjust the depth and complexity of explanations based on the perceived social context implied by honorifics.
    
    \item The responsiveness of AI models to Japanese honorifics highlights the importance of cultural elements in AI development for educational applications.
\end{enumerate}

This study reveals that the interaction between cultural linguistic elements and AI responses is a promising area for further exploration, with significant implications for the future of education and AI development.

\section*{Acknowledgements}

This research utilized Claude 3.5 Sonnet, an AI assistant developed by Anthropic, as a research support tool for manuscript preparation and discussion organization. To maintain objectivity and avoid potential conflicts of interest, Claude was intentionally excluded from the experimental data collection process, which involved other AI models.

\bibliographystyle{unsrtnat}
\bibliography{references}  

\begin{thebibliography}{4}
\providecommand{\natexlab}[1]{#1}
\providecommand{\url}[1]{\texttt{#1}}
\expandafter\ifx\csname urlstyle\endcsname\relax
  \providecommand{\doi}[1]{doi: #1}\else
  \providecommand{\doi}{doi: \begingroup \urlstyle{rm}\Url}\fi

\bibitem[Sarwari et~al.(2024)Sarwari, Javed, Mohd~Adnan, and Abdul~Wahab]{Sarwari2024-yf}
Abdul~Qahar Sarwari, Muhammad~Naeem Javed, Hamedi Mohd~Adnan, and Mohammad~Nubli Abdul~Wahab.
\newblock Assessment of the impacts of artificial intelligence ({AI}) on intercultural communication among postgraduate students in a multicultural university environment.
\newblock \emph{Sci. Rep.}, 14\penalty0 (1):\penalty0 13849, June 2024.
\newblock \doi{https://doi.org/10.1038/s41598-024-63276-5}.

\bibitem[Ge et~al.(2024)Ge, Xu, Misaki, Markus, and Tsai]{Ge2024-mx}
Xiao Ge, Chunchen Xu, Daigo Misaki, Hazel~Rose Markus, and Jeanne~L Tsai.
\newblock How culture shapes what people want from ai.
\newblock In \emph{Proceedings of the CHI Conference on Human Factors in Computing Systems}, CHI ’24. ACM, May 2024.
\newblock \doi{10.1145/3613904.3642660}.
\newblock URL \url{http://dx.doi.org/10.1145/3613904.3642660}.

\bibitem[Jeon(2022)]{Jeon2022-wv}
Yongwoog~Andrew Jeon.
\newblock Let me transfer you to our ai-based manager: Impact of manager-level job titles assigned to ai-based agents on marketing outcomes.
\newblock \emph{Journal of Business Research}, 145:\penalty0 892--904, 2022.
\newblock ISSN 0148-2963.
\newblock \doi{https://doi.org/10.1016/j.jbusres.2022.03.028}.
\newblock URL \url{https://www.sciencedirect.com/science/article/pii/S0148296322002569}.

\bibitem[Sato(2024)]{sato2024exploringeducationallandscapeai}
Keisuke Sato.
\newblock Exploring the educational landscape of ai: Large language models' approaches to explaining conservation of momentum in physics, 2024.
\newblock URL \url{https://arxiv.org/abs/2407.05308}.

\end{thebibliography}

\appendix 
\section*{Appendix}
The Japanese morphological analyser Janome was used for text pre-processing. A t-test was performed using the responses with "No address or name" as a baseline. A one-way analysis of variance was performed between all honorific titles.

\begin{table}[htbp]
\centering
\caption{Summary of Text Characteristics Across Different Models and Honorifics}
\label{tab:text_characteristics}
\resizebox{\textwidth}{!}{%
\begin{tabular}{llcccc}
\toprule
Model & Honorific & Character Count & Sentence Count & Avg. Sentence Length & Formality Ratio \\
\midrule
\multirow{13}{*}{G3} 
 & No Address & 274.78 (±66.07) & 5.52 (±1.46) & 50.45 (±6.78) & 1.00 (±0.00) \\
 & -chan (hiragana) & 276.40 (±48.40) & 5.44 (±1.17) & 51.90 (±9.08) & 1.00 (±0.00) \\
 & -kun (hiragana) & 274.48 (±60.59) & 5.24 (±1.21) & 52.90 (±7.00) & 1.00 (±0.02) \\
 & -kun & 305.74 (±50.57) & 5.86 (±1.17) & 52.89 (±6.86) & 1.00 (±0.00) \\
 & No honorific & 303.20 (±51.93) & 5.86 (±0.98) & 52.06 (±6.18) & 1.00 (±0.00) \\
 & -san (hiragana) & 293.88 (±45.76) & 5.86 (±1.23) & 51.41 (±9.32) & 1.00 (±0.00) \\
 & -senpai & 325.08 (±49.83) & 6.18 (±1.01) & 53.26 (±7.50) & 1.00 (±0.00) \\
 & -sama (hiragana) & 282.08 (±54.25) & 5.54 (±1.08) & 51.64 (±8.24) & 1.00 (±0.00) \\
 & -sama & 287.12 (±48.91) & 5.48 (±0.92) & 52.77 (±6.89) & 1.00 (±0.00) \\
 & -dono & 311.72 (±55.35) & 6.22 (±1.62) & 51.65 (±8.92) & 1.00 (±0.00) \\
 & -sensei & 305.86 (±46.11) & 5.84 (±1.10) & 53.28 (±7.84) & 1.00 (±0.00) \\
 & -hakase & 318.08 (±50.78) & 6.24 (±1.35) & 52.07 (±8.04) & 1.00 (±0.00) \\
 & -shisho & 327.76 (±52.47) & 6.54 (±1.04) & 50.52 (±6.33) & 1.00 (±0.00) \\
 & -Oshisho-sama & 298.68 (±56.89) & 5.70 (±1.25) & 53.29 (±7.40) & 1.00 (±0.00) \\
\midrule
\multirow{13}{*}{G4} 
 & No Address & 771.32 (±88.75) & 14.32 (±2.90) & 55.51 (±10.15) & 1.00 (±0.00) \\
 & -chan (hiragana) & 688.54 (±97.04) & 12.76 (±2.46) & 55.17 (±9.87) & 1.00 (±0.00) \\
 & -kun (hiragana) & 678.64 (±99.15) & 12.48 (±2.42) & 55.48 (±8.92) & 1.00 (±0.00) \\
 & -kun & 714.70 (±115.28) & 12.90 (±2.59) & 56.31 (±7.95) & 1.00 (±0.00) \\
 & No honorific & 712.58 (±102.90) & 12.66 (±2.55) & 57.52 (±9.01) & 1.00 (±0.00) \\
 & -san (hiragana) & 713.62 (±124.51) & 13.08 (±2.74) & 55.75 (±9.68) & 1.00 (±0.02) \\
 & -senpai & 671.16 (±116.77) & 12.50 (±2.51) & 54.41 (±7.15) & 1.00 (±0.00) \\
 & -sama (hiragana) & 734.54 (±112.63) & 13.06 (±2.35) & 57.34 (±10.44) & 1.00 (±0.00) \\
 & -sama & 737.70 (±85.28) & 13.26 (±2.43) & 56.73 (±7.88) & 1.00 (±0.00) \\
 & -dono & 683.18 (±106.01) & 12.30 (±1.99) & 55.97 (±6.58) & 1.00 (±0.01) \\
 & -sensei & 700.94 (±100.73) & 12.92 (±2.56) & 55.40 (±8.91) & 1.00 (±0.00) \\
 & -hakase & 751.52 (±113.82) & 13.42 (±2.34) & 56.99 (±9.45) & 1.00 (±0.00) \\
 & -shisho & 689.66 (±131.70) & 12.64 (±2.89) & 55.49 (±8.58) & 1.00 (±0.00) \\
 & -Oshisho-sama & 689.02 (±109.53) & 12.44 (±2.42) & 56.17 (±7.55) & 1.00 (±0.00) \\
\midrule
\multirow{13}{*}{C} 
 & No Address & 671.52 (±109.33) & 17.30 (±4.45) & 39.85 (±4.89) & 1.00 (±0.01) \\
 & -chan (hiragana) & 707.40 (±93.51) & 18.20 (±4.06) & 39.75 (±4.89) & 1.00 (±0.01) \\
 & -kun (hiragana) & 717.56 (±138.58) & 17.50 (±5.43) & 42.62 (±6.42) & 1.00 (±0.00) \\
 & -kun & 681.20 (±131.00) & 17.46 (±5.24) & 40.30 (±5.12) & 1.00 (±0.02) \\
 & No honorific & 776.62 (±86.90) & 20.74 (±4.26) & 38.49 (±6.12) & 1.00 (±0.00) \\
 & -san (hiragana) & 776.06 (±118.15) & 21.42 (±5.20) & 37.29 (±5.13) & 1.00 (±0.01) \\
 & -senpai & 682.62 (±126.81) & 16.88 (±3.63) & 40.89 (±4.11) & 1.00 (±0.02) \\
 & -sama (hiragana) & 748.32 (±87.46) & 19.66 (±4.38) & 39.22 (±5.86) & 1.00 (±0.00) \\
 & -sama & 703.94 (±112.83) & 17.40 (±4.46) & 41.62 (±5.79) & 1.00 (±0.00) \\
 & -dono & 707.88 (±102.51) & 18.80 (±4.68) & 38.86 (±5.73) & 1.00 (±0.01) \\
 & -sensei & 731.48 (±101.48) & 19.34 (±4.77) & 39.14 (±6.27) & 1.00 (±0.01) \\
 & -hakase & 758.66 (±97.09) & 20.88 (±4.55) & 37.31 (±5.64) & 1.00 (±0.01) \\
 & -shisho & 652.94 (±120.33) & 16.00 (±4.28) & 41.76 (±4.94) & 1.00 (±0.01) \\
 & -Oshisho-sama & 650.46 (±114.08) & 15.58 (±3.84) & 42.64 (±5.09) & 1.00 (±0.01) \\
\bottomrule
\end{tabular}%
}
\caption*{Note: Values are presented as mean (±standard deviation).}
\end{table}

\begin{table}[htbp]
\centering
\caption{Summary of Text Characteristics Across Different Models and Honorifics (Continued)}
\label{tab:text_characteristics_continued}
\resizebox{\textwidth}{!}{%
\begin{tabular}{llcccc}
\toprule
Model & Honorific & Character Count & Sentence Count & Avg. Sentence Length & Formality Ratio \\
\midrule
\multirow{13}{*}{Ge1p} 
 & No Address & 634.56 (±86.11) & 14.92 (±2.86) & 43.43 (±6.32) & 0.99 (±0.02) \\
 & -chan (hiragana) & 599.52 (±93.93) & 13.48 (±2.84) & 45.60 (±7.73) & 1.00 (±0.02) \\
 & -kun (hiragana) & 613.92 (±82.03) & 14.24 (±2.75) & 44.03 (±6.49) & 1.00 (±0.02) \\
 & -kun & 603.24 (±87.27) & 13.86 (±2.80) & 44.40 (±6.22) & 1.00 (±0.01) \\
 & No honorific & 567.60 (±78.69) & 13.54 (±3.08) & 43.05 (±6.03) & 0.99 (±0.02) \\
 & -san (hiragana) & 611.12 (±94.07) & 14.56 (±2.99) & 42.80 (±5.72) & 1.00 (±0.02) \\
 & -senpai & 619.58 (±97.63) & 13.70 (±3.01) & 46.13 (±5.84) & 1.00 (±0.02) \\
 & -sama (hiragana) & 626.68 (±87.53) & 14.92 (±3.06) & 43.06 (±6.75) & 0.99 (±0.03) \\
 & -sama & 629.34 (±74.44) & 14.78 (±2.62) & 43.26 (±5.15) & 1.00 (±0.01) \\
 & -dono & 633.12 (±98.20) & 14.66 (±2.76) & 43.75 (±5.25) & 1.00 (±0.01) \\
 & -sensei & 642.18 (±101.00) & 14.86 (±4.02) & 44.89 (±8.52) & 0.98 (±0.08) \\
 & -hakase & 646.14 (±91.27) & 14.94 (±2.95) & 44.06 (±5.98) & 0.99 (±0.02) \\
 & -shisho & 625.98 (±77.73) & 14.16 (±3.00) & 45.32 (±6.68) & 0.99 (±0.05) \\
 & -Oshisho-sama & 598.22 (±79.39) & 13.16 (±3.06) & 47.00 (±8.41) & 0.97 (±0.14) \\
\midrule
\multirow{13}{*}{Ge15f} 
 & No Address & 773.60 (±76.80) & 18.16 (±3.06) & 43.21 (±4.70) & 1.00 (±0.01) \\
 & -chan (hiragana) & 736.02 (±94.16) & 17.84 (±2.80) & 41.69 (±4.59) & 1.00 (±0.01) \\
 & -kun (hiragana) & 754.84 (±121.17) & 18.32 (±4.04) & 42.22 (±6.95) & 1.00 (±0.00) \\
 & -kun & 806.22 (±137.29) & 19.48 (±3.65) & 41.68 (±4.04) & 1.00 (±0.00) \\
 & No honorific & 712.94 (±90.39) & 16.68 (±2.72) & 43.29 (±5.56) & 1.00 (±0.00) \\
 & -san (hiragana) & 785.08 (±103.52) & 18.08 (±2.65) & 43.87 (±5.75) & 1.00 (±0.00) \\
 & -senpai & 793.66 (±127.67) & 19.54 (±3.66) & 41.09 (±4.87) & 1.00 (±0.01) \\
 & -sama (hiragana) & 807.54 (±101.53) & 18.80 (±3.06) & 43.56 (±5.77) & 0.99 (±0.02) \\
 & -sama & 798.92 (±115.65) & 18.58 (±3.40) & 43.60 (±5.49) & 1.00 (±0.00) \\
 & -dono & 739.18 (±127.83) & 17.42 (±3.52) & 42.87 (±4.99) & 1.00 (±0.01) \\
 & -sensei & 692.62 (±126.99) & 16.84 (±2.88) & 41.30 (±4.66) & 1.00 (±0.00) \\
 & -hakase & 678.26 (±160.26) & 16.74 (±3.90) & 40.76 (±4.98) & 1.00 (±0.00) \\
 & -shisho & 849.62 (±116.59) & 21.16 (±4.11) & 40.76 (±4.56) & 0.98 (±0.14) \\
 & -Oshisho-sama & 810.20 (±99.72) & 20.24 (±3.59) & 40.68 (±4.93) & 1.00 (±0.00) \\
\midrule
\multirow{13}{*}{Ge15p} 
 & No Address & 848.20 (±124.20) & 24.16 (±5.72) & 36.55 (±7.74) & 1.00 (±0.01) \\
 & -chan (hiragana) & 789.12 (±107.19) & 22.82 (±4.33) & 35.15 (±4.45) & 0.98 (±0.14) \\
 & -kun (hiragana) & 818.32 (±84.45) & 21.56 (±3.63) & 38.78 (±6.52) & 1.00 (±0.01) \\
 & -kun & 798.30 (±86.33) & 20.04 (±3.97) & 41.10 (±7.58) & 0.99 (±0.04) \\
 & No honorific & 861.00 (±139.09) & 20.52 (±4.65) & 43.00 (±6.79) & 0.99 (±0.02) \\
 & -san (hiragana) & 811.20 (±108.85) & 21.68 (±5.73) & 39.42 (±8.80) & 1.00 (±0.02) \\
 & -senpai & 800.80 (±117.49) & 22.26 (±3.75) & 36.52 (±5.52) & 0.82 (±0.38) \\
 & -sama (hiragana) & 840.52 (±102.90) & 21.06 (±4.20) & 40.91 (±6.61) & 0.99 (±0.03) \\
 & -sama & 830.98 (±100.10) & 20.92 (±4.62) & 41.19 (±8.09) & 0.99 (±0.04) \\
 & -dono & 818.72 (±98.04) & 21.22 (±3.99) & 39.33 (±5.30) & 0.97 (±0.14) \\
 & -sensei & 863.50 (±280.25) & 22.04 (±6.57) & 39.38 (±5.21) & 0.99 (±0.06) \\
 & -hakase & 963.24 (±152.37) & 23.58 (±5.27) & 41.67 (±5.53) & 1.00 (±0.01) \\
 & -shisho & 827.08 (±119.04) & 24.70 (±3.96) & 33.80 (±4.05) & 0.78 (±0.41) \\
 & -Oshisho-sama & 832.82 (±104.74) & 22.94 (±4.27) & 37.12 (±5.71) & 0.90 (±0.30) \\
\bottomrule
\end{tabular}%
}
\caption*{Note: Values are presented as mean (±standard deviation).}
\end{table}

\begin{table}[htbp]
\centering
\caption{t-test Results Across Models and Honorifics}
\label{tab:ttest_results_all}
\resizebox{\textwidth}{!}{%
\begin{tabular}{llccccc}
\toprule
Model & Honorific & Character Count & Sentence Count & Avg. Sentence Length & Polite Count & Physics Term Count \\
\midrule
\multirow{14}{*}{G3} 
 & No Address & - & - & - & - & - \\
 & -chan (hiragana) & - & - & - & - & - \\
 & -kun (hiragana) & - & - & - & - & - \\
 & -kun & \cellcolor{yellow!25}t=-2.60, p=0.011 & - & - & - & - \\
 & No honorific & - & - & - & - & - \\
 & -san (hiragana) & - & - & - & - & - \\
 & -senpai & \cellcolor{red!25}t=-4.25, p<0.001 & \cellcolor{yellow!25}t=-2.60, p=0.011 & - & \cellcolor{green!25}t=-3.25, p=0.002 & - \\
 & -sama (hiragana) & - & - & - & - & - \\
 & -sama & - & - & - & - & - \\
 & -dono & \cellcolor{green!25}t=-3.00, p=0.003 & \cellcolor{yellow!25}t=-2.25, p=0.027 & - & \cellcolor{yellow!25}t=-2.66, p=0.009 & - \\
 & -sensei & \cellcolor{yellow!25}t=-2.70, p=0.008 & - & - & \cellcolor{yellow!25}t=-2.10, p=0.038 & \cellcolor{green!25}t=-3.22, p=0.002 \\
 & -hakase & \cellcolor{red!25}t=-3.64, p<0.001 & \cellcolor{yellow!25}t=-2.54, p=0.013 & - & \cellcolor{green!25}t=-3.14, p=0.002 & \cellcolor{yellow!25}t=-2.55, p=0.012 \\
 & -shisho & \cellcolor{red!25}t=-4.40, p<0.001 & \cellcolor{red!25}t=-3.98, p<0.001 & - & \cellcolor{red!25}t=-4.67, p<0.001 & \cellcolor{yellow!25}t=-2.23, p=0.028 \\
 & -Oshisho-sama & - & - & - & - & - \\
\midrule
\multirow{14}{*}{G4}
 & No Address & - & - & - & - & - \\
 & -chan (hiragana) & \cellcolor{red!25}t=4.41, p<0.001 & \cellcolor{yellow!25}t=2.87, p=0.005 & - & \cellcolor{red!25}t=4.01, p<0.001 & \cellcolor{red!25}t=4.76, p<0.001 \\
 & -kun (hiragana) & \cellcolor{red!25}t=4.88, p<0.001 & \cellcolor{green!25}t=3.41, p<0.001 & - & \cellcolor{green!25}t=3.20, p=0.002 & \cellcolor{red!25}t=4.11, p<0.001 \\
 & -kun & \cellcolor{yellow!25}t=2.72, p=0.008 & \cellcolor{yellow!25}t=2.56, p=0.012 & - & \cellcolor{green!25}t=3.01, p=0.003 & \cellcolor{green!25}t=3.24, p=0.002 \\
 & No honorific & \cellcolor{yellow!25}t=2.78, p=0.007 & \cellcolor{yellow!25}t=2.97, p=0.004 & - & \cellcolor{yellow!25}t=2.04, p=0.045 & - \\
 & -san (hiragana) & \cellcolor{yellow!25}t=2.64, p=0.010 & \cellcolor{yellow!25}t=2.17, p=0.032 & - & \cellcolor{green!25}t=3.40, p=0.001 & \cellcolor{yellow!25}t=2.79, p=0.006 \\
 & -senpai & \cellcolor{red!25}t=4.78, p<0.001 & \cellcolor{green!25}t=3.32, p=0.001 & - & \cellcolor{green!25}t=3.10, p=0.003 & \cellcolor{green!25}t=3.36, p=0.001 \\
 & -sama (hiragana) & - & \cellcolor{yellow!25}t=2.36, p=0.020 & - & \cellcolor{green!25}t=3.42, p<0.001 & \cellcolor{yellow!25}t=2.79, p=0.006 \\
 & -sama & - & - & - & \cellcolor{yellow!25}t=2.04, p=0.045 & - \\
 & -dono & \cellcolor{red!25}t=4.46, p<0.001 & \cellcolor{red!25}t=4.02, p<0.001 & - & \cellcolor{red!25}t=3.76, p<0.001 & \cellcolor{red!25}t=4.69, p<0.001 \\
 & -sensei & \cellcolor{red!25}t=3.67, p<0.001 & \cellcolor{yellow!25}t=2.53, p=0.013 & - & \cellcolor{red!25}t=3.78, p<0.001 & \cellcolor{yellow!25}t=2.28, p=0.025 \\
 & -hakase & - & - & - & \cellcolor{yellow!25}t=2.06, p=0.042 & \cellcolor{yellow!25}t=2.91, p=0.005 \\
 & -shisho & \cellcolor{red!25}t=3.60, p<0.001 & \cellcolor{yellow!25}t=2.87, p=0.005 & - & \cellcolor{yellow!25}t=2.59, p=0.011 & \cellcolor{red!25}t=4.07, p<0.001 \\
 & -Oshisho-sama & \cellcolor{red!25}t=4.09, p<0.001 & \cellcolor{green!25}t=3.48, p<0.001 & - & \cellcolor{green!25}t=3.48, p<0.001 & \cellcolor{red!25}t=4.68, p<0.001 \\
\midrule
\multirow{14}{*}{C}
 & No Address & - & - & - & - & - \\
 & -chan (hiragana) & - & - & - & - & - \\
 & -kun (hiragana) & - & - & \cellcolor{yellow!25}t=-2.40, p=0.018 & - & \cellcolor{yellow!25}t=-2.80, p=0.006 \\
 & -kun & - & - & - & - & \cellcolor{yellow!25}t=-2.38, p=0.019 \\
 & No honorific & \cellcolor{red!25}t=-4.96, p<0.001 & \cellcolor{red!25}t=-3.85, p<0.001 & - & - & - \\
 & -san (hiragana) & \cellcolor{red!25}t=-4.55, p<0.001 & \cellcolor{red!25}t=-4.22, p<0.001 & \cellcolor{yellow!25}t=2.53, p=0.013 & \cellcolor{yellow!25}t=-2.97, p=0.004 & \cellcolor{green!25}t=-3.13, p=0.002 \\
 & -senpai & - & - & - & - & \cellcolor{yellow!25}t=-2.65, p=0.009 \\
 & -sama (hiragana) & \cellcolor{red!25}t=-3.84, p<0.001 & \cellcolor{yellow!25}t=-2.65, p=0.009 & - & - & \cellcolor{yellow!25}t=-2.61, p=0.011 \\
 & -sama & - & - & - & - & \cellcolor{yellow!25}t=-2.02, p=0.046 \\
 & -dono & - & - & - & - & \cellcolor{yellow!25}t=-2.95, p=0.004 \\
 & -sensei & \cellcolor{yellow!25}t=-2.81, p=0.006 & \cellcolor{yellow!25}t=-2.19, p=0.031 & - & - & - \\
 & -hakase & \cellcolor{red!25}t=-4.17, p<0.001 & \cellcolor{red!25}t=-3.94, p<0.001 & \cellcolor{yellow!25}t=2.39, p=0.019 & \cellcolor{yellow!25}t=-2.95, p=0.004 & \cellcolor{green!25}t=-3.06, p=0.003 \\
 & -shisho & - & - & - & \cellcolor{yellow!25}t=2.25, p=0.027 & - \\
 & -Oshisho-sama & - & \cellcolor{yellow!25}t=2.05, p=0.043 & \cellcolor{yellow!25}t=-2.76, p=0.007 & \cellcolor{yellow!25}t=2.86, p=0.005 & - \\
\bottomrule
\end{tabular}%
}
\caption*{Note: '-' indicates non-significant results (p $\geq$ 0.05). Color coding: \textcolor{red}{Red} (|t| > 3.5), \textcolor{green}{Green} (3 < |t| < 3.5), \textcolor{yellow}{Yellow} (2 < |t| < 3) indicate the strength of effects.}
\end{table}

\begin{table}[htbp]
\centering
\caption{t-test Results Across Models and Honorifics (Continued)}
\label{tab:ttest_results_all_continued}
\resizebox{\textwidth}{!}{%
\begin{tabular}{llccccc}
\toprule
Model & Honorific & Character Count & Sentence Count & Avg. Sentence Length & Polite Count & Physics Term Count \\
\midrule
\multirow{14}{*}{Ge1p} 
 & No Address & - & - & - & - & - \\
 & -chan (hiragana) & \cellcolor{yellow!25}t=1.92, p=0.057 & \cellcolor{yellow!25}t=2.50, p=0.014 & - & - & \cellcolor{yellow!25}t=2.65, p=0.009 \\
 & -kun (hiragana) & - & - & - & - & - \\
 & -kun & - & - & - & - & \cellcolor{yellow!25}t=2.93, p=0.004 \\
 & No honorific & \cellcolor{red!25}t=3.82, p<0.001 & \cellcolor{yellow!25}t=2.33, p=0.022 & - & \cellcolor{yellow!25}t=2.95, p=0.004 & \cellcolor{green!25}t=3.46, p<0.001 \\
 & -san (hiragana) & - & - & - & - & \cellcolor{yellow!25}t=1.98, p=0.050 \\
 & -senpai & - & \cellcolor{yellow!25}t=2.05, p=0.043 & \cellcolor{yellow!25}t=-2.19, p=0.031 & \cellcolor{yellow!25}t=2.08, p=0.040 & - \\
 & -sama (hiragana) & - & - & - & - & - \\
 & -sama & - & - & - & - & - \\
 & -dono & - & - & - & - & - \\
 & -sensei & - & - & - & - & - \\
 & -hakase & - & - & - & - & - \\
 & -shisho & - & - & - & \cellcolor{yellow!25}t=2.30, p=0.024 & \cellcolor{yellow!25}t=2.30, p=0.024 \\
 & -Oshisho-sama & \cellcolor{yellow!25}t=2.17, p=0.032 & \cellcolor{yellow!25}t=2.94, p=0.004 & \cellcolor{yellow!25}t=-2.37, p=0.020 & \cellcolor{yellow!25}t=2.35, p=0.021 & \cellcolor{red!25}t=3.60, p<0.001 \\
\midrule
\multirow{14}{*}{Ge15f}
 & No Address & - & - & - & - & - \\
 & -chan (hiragana) & \cellcolor{yellow!25}t=2.17, p=0.033 & - & - & - & \cellcolor{red!25}t=3.68, p<0.001 \\
 & -kun (hiragana) & - & - & - & - & \cellcolor{yellow!25}t=2.33, p=0.022 \\
 & -kun & - & - & - & - & - \\
 & No honorific & \cellcolor{red!25}t=3.56, p<0.001 & \cellcolor{yellow!25}t=2.60, p=0.011 & - & \cellcolor{yellow!25}t=2.64, p=0.010 & \cellcolor{red!25}t=4.51, p<0.001 \\
 & -san (hiragana) & - & - & - & - & - \\
 & -senpai & - & \cellcolor{yellow!25}t=-2.02, p=0.046 & \cellcolor{yellow!25}t=2.18, p=0.032 & - & - \\
 & -sama (hiragana) & \cellcolor{yellow!25}t=-1.87, p=0.065 & - & - & - & - \\
 & -sama & - & - & - & - & - \\
 & -dono & - & - & - & - & \cellcolor{green!25}t=3.04, p=0.003 \\
 & -sensei & \cellcolor{red!25}t=3.82, p<0.001 & \cellcolor{yellow!25}t=2.20, p=0.030 & \cellcolor{yellow!25}t=2.02, p=0.046 & - & \cellcolor{red!25}t=4.96, p<0.001 \\
 & -hakase & \cellcolor{red!25}t=3.76, p<0.001 & \cellcolor{yellow!25}t=2.00, p=0.048 & \cellcolor{yellow!25}t=2.50, p=0.014 & - & \cellcolor{red!25}t=5.16, p<0.001 \\
 & -shisho & \cellcolor{red!25}t=-3.81, p<0.001 & \cellcolor{red!25}t=-4.10, p<0.001 & \cellcolor{yellow!25}t=2.61, p=0.010 & - & - \\
 & -Oshisho-sama & \cellcolor{yellow!25}t=-2.04, p=0.045 & \cellcolor{green!25}t=-3.08, p=0.003 & \cellcolor{yellow!25}t=2.60, p=0.011 & - & - \\
\midrule
\multirow{14}{*}{Ge15p}
 & No Address & - & - & - & - & - \\
 & -chan (hiragana) & \cellcolor{yellow!25}t=2.52, p=0.013 & - & - & - & \cellcolor{yellow!25}t=2.43, p=0.017 \\
 & -kun (hiragana) & - & \cellcolor{yellow!25}t=2.69, p=0.009 & - & - & - \\
 & -kun & \cellcolor{yellow!25}t=2.31, p=0.023 & \cellcolor{red!25}t=4.14, p<0.001 & \cellcolor{yellow!25}t=-2.94, p=0.004 & \cellcolor{yellow!25}t=2.21, p=0.030 & - \\
 & No honorific & - & \cellcolor{green!25}t=3.30, p=0.001 & \cellcolor{red!25}t=-4.05, p<0.001 & - & - \\
 & -san (hiragana) & - & \cellcolor{yellow!25}t=2.14, p=0.035 & - & - & - \\
 & -senpai & - & - & - & \cellcolor{yellow!25}t=2.92, p=0.004 & - \\
 & -sama (hiragana) & - & \cellcolor{green!25}t=3.06, p=0.003 & \cellcolor{green!25}t=-3.00, p=0.004 & - & - \\
 & -sama & - & \cellcolor{green!25}t=3.08, p=0.003 & \cellcolor{yellow!25}t=-2.90, p=0.005 & - & - \\
 & -dono & - & \cellcolor{yellow!25}t=2.95, p=0.004 & \cellcolor{yellow!25}t=-2.07, p=0.041 & - & - \\
 & -sensei & - & - & \cellcolor{yellow!25}t=-2.12, p=0.037 & \cellcolor{red!25}t=-3.66, p<0.001 & \cellcolor{yellow!25}t=2.54, p=0.013 \\
 & -hakase & \cellcolor{red!25}t=-4.10, p<0.001 & - & \cellcolor{red!25}t=-3.76, p<0.001 & \cellcolor{red!25}t=-7.64, p<0.001 & - \\
 & -shisho & - & - & \cellcolor{yellow!25}t=2.20, p=0.030 & \cellcolor{yellow!25}t=2.75, p=0.007 & - \\
 & -Oshisho-sama & - & - & - & \cellcolor{yellow!25}t=1.89, p=0.062 & \cellcolor{yellow!25}t=2.34, p=0.021 \\
\bottomrule
\end{tabular}%
}
\caption*{Note: '-' indicates non-significant results (p $\geq$ 0.05). Color coding: \textcolor{red}{Red} (|t| > 3.5), \textcolor{green}{Green} (3 < |t| < 3.5), \textcolor{yellow}{Yellow} (2 < |t| < 3) indicate the strength of effects.}
\end{table}

\begin{table}[htbp]
\centering
\caption{ANOVA Results by Model}
\label{tab:anova-models}
\begin{tabular}{lS[table-format=1.2]cS[table-format=1.2]cS[table-format=1.2]cS[table-format=2.2]c}
\toprule
Model & {Equation} & {Sig.} & {Char} & {Sig.} & {Sentence} & {Sig.} & {Avg Sent} & {Sig.} \\
      & {Count} &  & {Count} &  & {Count} &  & {Length} &  \\
\midrule
ChatGPT 3.5    & 2.96 & {***} & 5.77 & {***} & 4.63 & {***} & 0.77 &       \\
ChatGPT 4.0    & 1.20 &       & 3.71 & {***} & 2.09 & {*}   & 0.49 &       \\
Command R+     & 3.01 & {***} & 7.03 & {***} & 7.91 & {***} & 5.05 & {***} \\
Gemini 1.0 pro  & 0.96 &       & 2.76 & {***} & 2.10 & {*}   & 1.85 & {*}   \\
Gemini 1.5 flash & 2.33 & {**}  & 8.99 & {***} & 7.60 & {***} & 2.57 & {**}  \\
Gemini 1.5 pro & 2.63 & {**}  & 5.19 & {***} & 4.26 & {***} & 8.56 & {***} \\
\bottomrule
\multicolumn{9}{l}{\scriptsize{Significance levels: * p < 0.05, ** p < 0.01, *** p < 0.001}}
\end{tabular}
\end{table}

\begin{table}[htbp]
\centering
\caption{Similarity Metrics for Models G3, G4, and C}
\label{tab:similarity-g3-g4-c-all}
\begin{tabular}{lcccccc}
\toprule
\multirow{2}{*}{Honorific} & \multicolumn{2}{c}{Model G3} & \multicolumn{2}{c}{Model G4} & \multicolumn{2}{c}{Model C} \\
\cmidrule(lr){2-3} \cmidrule(lr){4-5} \cmidrule(lr){6-7}
& Cosine & LSA & Cosine & LSA & Cosine & LSA \\
\midrule
No Address & \cellcolor{blue!5}0.5655 & \cellcolor{blue!15}0.7304 & \cellcolor{blue!25}0.7552 & \cellcolor{blue!35}0.8576 & \cellcolor{blue!35}0.8447 & \cellcolor{blue!35}0.9222 \\
-chan (hiragana) & \cellcolor{blue!5}0.5960 & \cellcolor{blue!25}0.7893 & \cellcolor{blue!15}0.7136 & \cellcolor{blue!25}0.8311 & \cellcolor{blue!25}0.8180 & \cellcolor{blue!35}0.9083 \\
-kun (hiragana) & \cellcolor{blue!5}0.5801 & \cellcolor{blue!25}0.7683 & \cellcolor{blue!15}0.7238 & \cellcolor{blue!25}0.8376 & \cellcolor{blue!25}0.7905 & \cellcolor{blue!35}0.8931 \\
-kun & \cellcolor{blue!15}0.6346 & \cellcolor{blue!25}0.8226 & \cellcolor{blue!15}0.7239 & \cellcolor{blue!25}0.8285 & \cellcolor{blue!25}0.7914 & \cellcolor{blue!35}0.8866 \\
No honorific & \cellcolor{blue!5}0.5921 & \cellcolor{blue!25}0.7714 & \cellcolor{blue!15}0.7329 & \cellcolor{blue!35}0.8478 & \cellcolor{blue!35}0.8514 & \cellcolor{blue!35}0.9240 \\
-san (hiragana) & \cellcolor{blue!15}0.6122 & \cellcolor{blue!25}0.8021 & \cellcolor{blue!15}0.7136 & \cellcolor{blue!25}0.8253 & \cellcolor{blue!35}0.8332 & \cellcolor{blue!35}0.9163 \\
-senpai & \cellcolor{blue!5}0.5689 & \cellcolor{blue!15}0.7449 & \cellcolor{blue!15}0.7188 & \cellcolor{blue!35}0.8421 & \cellcolor{blue!25}0.7582 & \cellcolor{blue!35}0.8599 \\
-sama (hiragana) & \cellcolor{blue!5}0.5905 & \cellcolor{blue!25}0.7732 & \cellcolor{blue!15}0.7278 & \cellcolor{blue!25}0.8326 & \cellcolor{blue!35}0.8289 & \cellcolor{blue!35}0.9154 \\
-sama & \cellcolor{blue!5}0.5832 & \cellcolor{blue!25}0.7576 & \cellcolor{blue!15}0.7488 & \cellcolor{blue!35}0.8635 & \cellcolor{blue!25}0.8174 & \cellcolor{blue!35}0.9037 \\
-dono & \cellcolor{blue!5}0.5682 & \cellcolor{blue!15}0.7496 & \cellcolor{blue!15}0.7177 & \cellcolor{blue!25}0.8310 & \cellcolor{blue!25}0.8132 & \cellcolor{blue!35}0.9021 \\
-sensei & \cellcolor{blue!15}0.6108 & \cellcolor{blue!25}0.8004 & \cellcolor{blue!15}0.7253 & \cellcolor{blue!35}0.8390 & \cellcolor{blue!25}0.8182 & \cellcolor{blue!35}0.9076 \\
-hakase & \cellcolor{blue!5}0.5936 & \cellcolor{blue!25}0.7604 & \cellcolor{blue!15}0.7290 & \cellcolor{blue!35}0.8422 & \cellcolor{blue!35}0.8303 & \cellcolor{blue!35}0.9112 \\
-shisho & \cellcolor{blue!5}0.5528 & \cellcolor{blue!15}0.7127 & \cellcolor{blue!15}0.7308 & \cellcolor{blue!35}0.8519 & \cellcolor{blue!25}0.7707 & \cellcolor{blue!35}0.8744 \\
-Oshisho-sama & \cellcolor{blue!5}0.5770 & \cellcolor{blue!15}0.7542 & \cellcolor{blue!15}0.7149 & \cellcolor{blue!25}0.8274 & \cellcolor{blue!25}0.7821 & \cellcolor{blue!35}0.8787 \\
\bottomrule
\end{tabular}
\end{table}

\begin{table}
\centering
\small
\begin{tabular}{ccccc}
\cellcolor{blue!5} 0.55-0.65 &
\cellcolor{blue!15} 0.65-0.75 &
\cellcolor{blue!25} 0.75-0.85 &
\cellcolor{blue!35} 0.85-0.95 &
\cellcolor{blue!45} 0.95+ \\
\end{tabular}
\caption*{Color legend for similarity values}
\end{table}

\begin{table}[htbp]
\centering
\caption{Similarity Metrics for Models Ge1p, Ge15f, and Ge15p}
\label{tab:similarity-ge1p-ge15f-ge15p-all}
\begin{tabular}{lcccccc}
\toprule
\multirow{2}{*}{Honorific} & \multicolumn{2}{c}{Model Ge1p} & \multicolumn{2}{c}{Model Ge15f} & \multicolumn{2}{c}{Model Ge15p} \\
\cmidrule(lr){2-3} \cmidrule(lr){4-5} \cmidrule(lr){6-7}
& Cosine & LSA & Cosine & LSA & Cosine & LSA \\
\midrule
No Address & \cellcolor{blue!25}0.7581 & \cellcolor{blue!35}0.8731 & \cellcolor{blue!25}0.8152 & \cellcolor{blue!35}0.9029 & \cellcolor{blue!15}0.7167 & \cellcolor{blue!25}0.8306 \\
-chan (hiragana) & \cellcolor{blue!15}0.7242 & \cellcolor{blue!35}0.8562 & \cellcolor{blue!25}0.7579 & \cellcolor{blue!35}0.8772 & \cellcolor{blue!15}0.7398 & \cellcolor{blue!35}0.8535 \\
-kun (hiragana) & \cellcolor{blue!15}0.7159 & \cellcolor{blue!35}0.8536 & \cellcolor{blue!25}0.7624 & \cellcolor{blue!35}0.8711 & \cellcolor{blue!15}0.7143 & \cellcolor{blue!25}0.8280 \\
-kun & \cellcolor{blue!15}0.7199 & \cellcolor{blue!35}0.8499 & \cellcolor{blue!25}0.7640 & \cellcolor{blue!35}0.8741 & \cellcolor{blue!25}0.7699 & \cellcolor{blue!35}0.8756 \\
No honorific & \cellcolor{blue!15}0.7273 & \cellcolor{blue!35}0.8547 & \cellcolor{blue!25}0.7701 & \cellcolor{blue!35}0.8660 & \cellcolor{blue!15}0.7354 & \cellcolor{blue!35}0.8390 \\
-san (hiragana) & \cellcolor{blue!15}0.7316 & \cellcolor{blue!35}0.8496 & \cellcolor{blue!25}0.7866 & \cellcolor{blue!35}0.8933 & \cellcolor{blue!15}0.7329 & \cellcolor{blue!35}0.8449 \\
-senpai & \cellcolor{blue!15}0.7344 & \cellcolor{blue!35}0.8700 & \cellcolor{blue!25}0.7853 & \cellcolor{blue!35}0.8990 & \cellcolor{blue!15}0.7184 & \cellcolor{blue!35}0.8485 \\
-sama (hiragana) & \cellcolor{blue!15}0.7357 & \cellcolor{blue!35}0.8634 & \cellcolor{blue!25}0.7648 & \cellcolor{blue!35}0.8747 & \cellcolor{blue!15}0.7315 & \cellcolor{blue!35}0.8440 \\
-sama & \cellcolor{blue!15}0.7441 & \cellcolor{blue!35}0.8661 & \cellcolor{blue!25}0.8011 & \cellcolor{blue!35}0.9007 & \cellcolor{blue!25}0.7616 & \cellcolor{blue!35}0.8731 \\
-dono & \cellcolor{blue!15}0.7471 & \cellcolor{blue!35}0.8666 & \cellcolor{blue!25}0.7737 & \cellcolor{blue!35}0.8796 & \cellcolor{blue!15}0.7257 & \cellcolor{blue!25}0.8310 \\
-sensei & \cellcolor{blue!15}0.7403 & \cellcolor{blue!35}0.8732 & \cellcolor{blue!25}0.7589 & \cellcolor{blue!35}0.8719 & \cellcolor{blue!15}0.6970 & \cellcolor{blue!25}0.8128 \\
-hakase & \cellcolor{blue!25}0.7553 & \cellcolor{blue!35}0.8840 & \cellcolor{blue!25}0.7509 & \cellcolor{blue!35}0.8638 & \cellcolor{blue!15}0.7349 & \cellcolor{blue!25}0.8245 \\
-shisho & \cellcolor{blue!15}0.7221 & \cellcolor{blue!35}0.8541 & \cellcolor{blue!25}0.7874 & \cellcolor{blue!35}0.8994 & \cellcolor{blue!15}0.7310 & \cellcolor{blue!35}0.8503 \\
-Oshisho-sama & \cellcolor{blue!15}0.6961 & \cellcolor{blue!25}0.8364 & \cellcolor{blue!25}0.7817 & \cellcolor{blue!35}0.8922 & \cellcolor{blue!15}0.7124 & \cellcolor{blue!35}0.8404 \\
\bottomrule
\end{tabular}
\end{table}

\begin{table}
\centering
\small
\begin{tabular}{ccccc}
\cellcolor{blue!5} 0.55-0.65 &
\cellcolor{blue!15} 0.65-0.75 &
\cellcolor{blue!25} 0.75-0.85 &
\cellcolor{blue!35} 0.85-0.95 &
\cellcolor{blue!45} 0.95+ \\
\end{tabular}
\caption*{Color legend for similarity values}
\end{table}

\begin{table}[htbp]
\centering
\caption{t-test Results for Similarity Metrics Across All Models and Honorifics}
\label{tab:all_similarity_ttest_highlighted}
\resizebox{\textwidth}{!}{%
\begin{tabular}{l|cc|cc|cc}
\toprule
\multirow{2}{*}{Honorific} & \multicolumn{2}{c|}{Model G3} & \multicolumn{2}{c|}{Model G4} & \multicolumn{2}{c|}{Model C} \\
\cmidrule{2-7}
 & Cosine & LSA & Cosine & LSA & Cosine & LSA \\
\midrule
No Address & \cellcolor{yellow!25}-2.17* & \cellcolor{yellow!25}-2.32* & \cellcolor{green!25}2.78** & 1.01 & -1.31 & -0.32 \\
-chan (hiragana) & \cellcolor{yellow!25}-2.64* & \cellcolor{red!25}-3.33** & \cellcolor{green!25}5.02*** & \cellcolor{yellow!25}2.53* & \cellcolor{blue!25}5.05*** & \cellcolor{yellow!25}2.26* \\
-kun (hiragana) & -1.41 & \cellcolor{yellow!25}-2.49* & \cellcolor{green!25}3.85*** & \cellcolor{yellow!25}2.05* & \cellcolor{blue!25}8.19*** & \cellcolor{blue!25}3.71*** \\
-kun & \cellcolor{red!25}-6.54*** & \cellcolor{red!25}-5.77*** & \cellcolor{green!25}3.99*** & \cellcolor{green!25}3.18** & \cellcolor{blue!25}11.28*** & \cellcolor{blue!25}6.80*** \\
No honorific & \cellcolor{yellow!25}-2.17* & \cellcolor{yellow!25}-2.32* & \cellcolor{green!25}2.78** & 1.01 & -1.31 & -0.32 \\
-san (hiragana) & \cellcolor{red!25}-4.14*** & \cellcolor{red!25}-4.31*** & \cellcolor{green!25}4.90*** & \cellcolor{green!25}3.23** & \cellcolor{green!25}2.92** & 1.28 \\
-senpai & -0.30 & -0.83 & \cellcolor{green!25}4.65*** & 1.65 & \cellcolor{blue!25}8.53*** & \cellcolor{blue!25}5.13*** \\
-sama (hiragana) & \cellcolor{yellow!25}-2.10* & \cellcolor{yellow!25}-2.45* & \cellcolor{green!25}3.40** & \cellcolor{green!25}2.66** & \cellcolor{green!25}3.29** & 1.24 \\
-sama & -1.41 & -1.47 & 0.90 & -0.68 & \cellcolor{blue!25}5.51*** & \cellcolor{green!25}3.15** \\
-dono & -0.17 & -0.83 & \cellcolor{green!25}4.82*** & \cellcolor{green!25}2.88** & \cellcolor{blue!25}7.85*** & \cellcolor{blue!25}4.46*** \\
-sensei & \cellcolor{red!25}-3.75*** & \cellcolor{red!25}-3.88*** & \cellcolor{green!25}4.06*** & \cellcolor{yellow!25}2.09* & \cellcolor{blue!25}5.30*** & \cellcolor{yellow!25}2.54* \\
-hakase & \cellcolor{yellow!25}-2.19* & -1.57 & \cellcolor{green!25}3.21** & 1.56 & \cellcolor{green!25}2.94** & 1.88 \\
-shisho & 0.86 & 0.83 & \cellcolor{green!25}3.25** & 0.63 & \cellcolor{blue!25}10.15*** & \cellcolor{blue!25}5.53*** \\
-Oshisho-sama & -0.83 & -1.20 & \cellcolor{green!25}4.89*** & \cellcolor{green!25}3.06** & \cellcolor{blue!25}10.29*** & \cellcolor{blue!25}6.30*** \\
\midrule
 & \multicolumn{2}{c|}{Model Ge1p} & \multicolumn{2}{c|}{Model Ge15f} & \multicolumn{2}{c}{Model Ge15p} \\
\cmidrule{2-7}
 & Cosine & LSA & Cosine & LSA & Cosine & LSA \\
\midrule
No Address & \cellcolor{yellow!25}2.65* & 1.23 & \cellcolor{blue!25}5.27*** & \cellcolor{green!25}3.55** & -1.18 & -0.42 \\
-chan (hiragana) & \cellcolor{green!25}3.41** & 1.31 & \cellcolor{blue!25}6.23*** & \cellcolor{yellow!25}2.25* & -1.66 & -1.31 \\
-kun (hiragana) & \cellcolor{green!25}4.32*** & 1.53 & \cellcolor{blue!25}6.47*** & \cellcolor{green!25}3.19** & 0.14 & 0.12 \\
-kun & \cellcolor{green!25}4.40*** & \cellcolor{yellow!25}2.07* & \cellcolor{blue!25}6.40*** & \cellcolor{green!25}2.97** & \cellcolor{red!25}-3.85*** & \cellcolor{yellow!25}-2.60* \\
No honorific & \cellcolor{yellow!25}2.65* & 1.23 & \cellcolor{blue!25}5.27*** & \cellcolor{green!25}3.55** & -1.18 & -0.42 \\
-san (hiragana) & \cellcolor{green!25}2.91** & \cellcolor{yellow!25}2.01* & \cellcolor{green!25}4.37*** & 1.19 & -1.04 & -0.74 \\
-senpai & \cellcolor{green!25}3.00** & 0.31 & \cellcolor{green!25}4.36*** & 0.48 & -0.12 & -0.94 \\
-sama (hiragana) & \cellcolor{green!25}2.66** & 0.90 & \cellcolor{blue!25}6.49*** & \cellcolor{green!25}2.89** & -0.91 & -0.66 \\
-sama & 1.65 & 0.65 & \cellcolor{yellow!25}2.30* & 0.30 & \cellcolor{red!25}-3.02** & \cellcolor{yellow!25}-2.25* \\
-dono & 1.39 & 0.63 & \cellcolor{blue!25}5.13*** & \cellcolor{yellow!25}2.35* & -0.55 & -0.02 \\
-sensei & 1.95 & -0.01 & \cellcolor{blue!25}7.93*** & \cellcolor{green!25}3.49** & 1.19 & 0.84 \\
-hakase & 0.35 & -1.03 & \cellcolor{blue!25}8.58*** & \cellcolor{green!25}4.35*** & -1.00 & 0.27 \\
-shisho & \cellcolor{green!25}3.48** & 1.42 & \cellcolor{green!25}3.87*** & 0.40 & -1.04 & -1.15 \\
-Oshisho-sama & \cellcolor{blue!25}5.70*** & \cellcolor{yellow!25}2.56* & \cellcolor{green!25}4.58*** & 1.20 & 0.26 & -0.47 \\
\bottomrule
\end{tabular}%
}
\caption*{Note: Values are t-statistics. Significance levels: * p < 0.05, ** p < 0.01, *** p < 0.001. 
Color coding: \textcolor{red}{Red} (t < -3), \textcolor{yellow}{Yellow} (-3 < t < -2 or 2 < t < 3), 
\textcolor{green}{Green} (3 < t < 5), \textcolor{blue}{Blue} (t > 5) indicate the strength and direction of effects.}
\end{table}

\begin{table}[htbp]
\centering
\caption{One-way ANOVA Results for Cosine Similarity and LSA}
\label{tab:anova-results-combined_cosineLSA}
\begin{tabular}{lcccc}
\toprule
\multirow{2}{*}{Model} & \multicolumn{2}{c}{Cosine Similarity} & \multicolumn{2}{c}{LSA} \\
\cmidrule(lr){2-3} \cmidrule(lr){4-5}
 & F value & p value & F value & p value \\
\midrule
ChatGPT 3.5    & 4.593  & 1.348e-07*** & 4.166  & 1.000e-06*** \\
ChatGPT 4.0    & 5.072  & 1.252e-08*** & 3.140  & 1.400e-04*** \\
Command R+     & 29.750 & 3.032e-58*** & 10.013 & 1.895e-19*** \\
Gemini 1.0 pro  & 5.793  & 3.395e-10*** & 1.900  & 2.718e-02*   \\
Gemini 1.5 flash & 10.416 & 2.524e-20*** & 4.008  & 2.000e-06*** \\
Gemini 1.5 pro & 3.465  & 3.100e-05*** & 1.839  & 3.416e-02*   \\
\bottomrule
\multicolumn{5}{l}{\footnotesize{Significance levels: * p < 0.05, ** p < 0.01, *** p < 0.001}}
\end{tabular}
\end{table}

\begin{table}
\centering
\begin{tabular}{lcccccccc}
\toprule
Honorific & Vector & Time & Momentum & Net force & Quantum  & Derivation & Inelastic  & Conserv. energy \\
\midrule
No Address & 4 & 3 & 0 & 4 & 2 & 0 & 0 & 3 \\
-chan (hiragana) & 8 & 6 & 0 & 3 & 7 & 0 & 0 & 13 \\
-kun (hiragana) & 2 & 3 & 0 & 1 & 2 & 1 & 1 & 13 \\
-kun & 8 & 4 & 0 & 4 & 2 & 1 & 0 & 12 \\
No honorific & 7 & 7 & 0 & 3 & 3 & 0 & 1 & 18 \\
-san (hiragana) & 6 & 5 & 0 & 5 & 2 & 1 & 1 & 14 \\
-senpai & 4 & 5 & 0 & 5 & 0 & 0 & 1 & 14 \\
-sama (hiragana) & 7 & 5 & 0 & 4 & 4 & 1 & 1 & 8 \\
-sama & 4 & 12 & 0 & 3 & 4 & 0 & 1 & 5 \\
-dono & 7 & 5 & 1 & 1 & 4 & 0 & 2 & 9 \\
-sensei & 5 & 8 & 0 & 3 & 3 & 0 & 0 & 20 \\
-hakase & 3 & 12 & 0 & 6 & 4 & 1 & 0 & 10 \\
-shisho & 5 & 5 & 0 & 7 & 2 & 2 & 2 & 13 \\
-Oshisho-sama & 7 & 7 & 0 & 2 & 5 & 1 & 1 & 8 \\
\bottomrule
\end{tabular}
\caption{Model ChatGPT 3.5 Keyword Search Results}
\label{tab:model-g3-keywords}
\end{table}

\begin{table}
\centering
\begin{tabular}{lcccccccc}
\toprule
Honorific & Vector & Time & Momentum & Net force & Quantum  & Derivation & Inelastic  & Conserv. energy \\
\midrule
No Address & 48 & 31 & 0 & 14 & 7 & 4 & 15 & 9 \\
-chan (hiragana) & 40 & 25 & 0 & 16 & 5 & 6 & 8 & 8 \\
-kun (hiragana) & 41 & 30 & 0 & 9 & 5 & 3 & 11 & 7 \\
-kun & 41 & 31 & 0 & 19 & 8 & 5 & 8 & 9 \\
No honorific & 46 & 31 & 1 & 21 & 8 & 3 & 12 & 13 \\
-san (hiragana) & 42 & 40 & 0 & 12 & 4 & 4 & 15 & 7 \\
-senpai & 43 & 30 & 0 & 20 & 6 & 3 & 9 & 12 \\
-sama (hiragana) & 41 & 29 & 0 & 18 & 5 & 0 & 11 & 11 \\
-sama & 46 & 34 & 0 & 20 & 11 & 3 & 14 & 10 \\
-dono & 45 & 27 & 0 & 18 & 8 & 1 & 10 & 10 \\
-sensei & 44 & 28 & 0 & 21 & 9 & 1 & 14 & 6 \\
-hakase & 43 & 30 & 0 & 14 & 5 & 3 & 16 & 10 \\
-shisho & 45 & 31 & 1 & 14 & 2 & 3 & 8 & 7 \\
-Oshisho-sama & 42 & 29 & 0 & 11 & 5 & 1 & 13 & 9 \\
\bottomrule
\end{tabular}
\caption{Model G4 Keyword Search Results}
\label{tab:model-g4-keywords}
\end{table}

\begin{table}
\centering
\begin{tabular}{lcccccccc}
\toprule
Honorific & Vector & Time & Momentum & Net force & Quantum  & Derivation & Inelastic  & Conserv. energy \\
\midrule
No Address & 31 & 42 & 0 & 3 & 21 & 18 & 30 & 9 \\
-chan (hiragana) & 32 & 48 & 0 & 6 & 15 & 17 & 25 & 14 \\
-kun (hiragana) & 28 & 44 & 0 & 14 & 14 & 28 & 26 & 19 \\
-kun & 32 & 48 & 0 & 8 & 19 & 34 & 26 & 15 \\
No honorific & 32 & 44 & 0 & 8 & 16 & 18 & 30 & 4 \\
-san (hiragana) & 34 & 49 & 0 & 10 & 14 & 14 & 31 & 9 \\
-senpai & 26 & 41 & 0 & 12 & 24 & 23 & 25 & 27 \\
-sama (hiragana) & 36 & 46 & 0 & 7 & 15 & 18 & 36 & 11 \\
-sama & 30 & 47 & 0 & 12 & 15 & 24 & 31 & 9 \\
-dono & 28 & 49 & 0 & 16 & 18 & 17 & 20 & 16 \\
-sensei & 40 & 43 & 0 & 7 & 17 & 19 & 27 & 8 \\
-hakase & 34 & 39 & 0 & 6 & 13 & 19 & 32 & 9 \\
-shisho & 18 & 48 & 0 & 9 & 17 & 26 & 21 & 23 \\
-Oshisho-sama & 24 & 45 & 0 & 15 & 31 & 22 & 25 & 20 \\
\bottomrule
\end{tabular}
\caption{Model C Keyword Search Results}
\label{tab:model-c-keywords}
\end{table}

\begin{table}
\centering
\begin{tabular}{lcccccccc}
\toprule
Honorific & Vector & Time & Momentum & Net force & Quantum  & Derivation & Inelastic  & Conserv. energy \\
\midrule
No Address & 19 & 41 & 0 & 0 & 3 & 4 & 8 & 6 \\
-chan (hiragana) & 23 & 42 & 0 & 1 & 1 & 5 & 8 & 1 \\
-kun (hiragana) & 28 & 43 & 0 & 0 & 4 & 4 & 5 & 6 \\
-kun & 25 & 44 & 0 & 3 & 2 & 4 & 11 & 5 \\
No honorific & 22 & 39 & 0 & 2 & 1 & 0 & 5 & 1 \\
-san (hiragana) & 32 & 37 & 0 & 0 & 1 & 4 & 6 & 0 \\
-senpai & 30 & 44 & 0 & 5 & 5 & 8 & 10 & 4 \\
-sama (hiragana) & 29 & 41 & 0 & 2 & 2 & 6 & 8 & 4 \\
-sama & 27 & 42 & 0 & 2 & 3 & 8 & 9 & 3 \\
-dono & 26 & 43 & 0 & 3 & 0 & 7 & 8 & 3 \\
-sensei & 20 & 38 & 0 & 5 & 2 & 4 & 11 & 1 \\
-hakase & 27 & 44 & 0 & 1 & 4 & 3 & 6 & 5 \\
-shisho & 25 & 42 & 0 & 2 & 1 & 6 & 18 & 5 \\
-Oshisho-sama & 28 & 46 & 0 & 2 & 3 & 5 & 6 & 1 \\
\bottomrule
\end{tabular}
\caption{Model Ge1p Keyword Search Results}
\label{tab:model-ge1p-keywords}
\end{table}

\begin{table}
\centering
\begin{tabular}{lcccccccc}
\toprule
Honorific & Vector & Time & Momentum & Net force & Quantum  & Derivation & Inelastic  & Conserv. energy \\
\midrule
No Address & 36 & 6 & 2 & 2 & 1 & 9 & 0 & 2 \\
-chan (hiragana) & 29 & 0 & 2 & 1 & 0 & 6 & 1 & 4 \\
-kun (hiragana) & 32 & 4 & 4 & 1 & 0 & 13 & 3 & 3 \\
-kun & 30 & 3 & 5 & 4 & 1 & 11 & 4 & 7 \\
No honorific & 34 & 2 & 1 & 0 & 1 & 13 & 2 & 2 \\
-san (hiragana) & 39 & 3 & 2 & 6 & 0 & 14 & 4 & 3 \\
-senpai & 32 & 5 & 1 & 1 & 0 & 13 & 4 & 2 \\
-sama (hiragana) & 35 & 6 & 3 & 1 & 0 & 25 & 4 & 5 \\
-sama & 31 & 5 & 2 & 0 & 0 & 17 & 2 & 9 \\
-dono & 32 & 6 & 3 & 4 & 1 & 13 & 0 & 4 \\
-sensei & 30 & 2 & 4 & 1 & 1 & 9 & 4 & 3 \\
-hakase & 28 & 5 & 0 & 0 & 1 & 17 & 2 & 4 \\
-shisho & 33 & 6 & 2 & 0 & 1 & 13 & 1 & 6 \\
-Oshisho-sama & 37 & 4 & 0 & 2 & 0 & 17 & 4 & 3 \\
\bottomrule
\end{tabular}
\caption{Model Ge15f Keyword Search Results}
\label{tab:model-ge15f-keywords}
\end{table}

\begin{table}
\centering
\begin{tabular}{lcccccccc}
\toprule
Honorific & Vector & Time & Momentum & Net force & Quantum  & Derivation & Inelastic  & Conserv. energy \\
\midrule
No Address & 44 & 5 & 9 & 0 & 4 & 3 & 1 & 5 \\
-chan (hiragana) & 39 & 12 & 10 & 2 & 0 & 3 & 1 & 6 \\
-kun (hiragana) & 43 & 8 & 11 & 1 & 2 & 4 & 6 & 7 \\
-kun & 37 & 18 & 19 & 1 & 3 & 7 & 2 & 11 \\
No honorific & 39 & 9 & 6 & 2 & 2 & 1 & 1 & 6 \\
-san (hiragana) & 34 & 15 & 8 & 1 & 4 & 7 & 3 & 7 \\
-senpai & 36 & 11 & 14 & 1 & 1 & 0 & 1 & 6 \\
-sama (hiragana) & 41 & 11 & 10 & 1 & 6 & 5 & 1 & 6 \\
-sama & 36 & 11 & 10 & 1 & 5 & 1 & 1 & 6 \\
-dono & 43 & 7 & 10 & 0 & 2 & 4 & 3 & 5 \\
-sensei & 39 & 10 & 2 & 3 & 4 & 6 & 4 & 2 \\
-hakase & 44 & 11 & 0 & 6 & 4 & 3 & 1 & 2 \\
-shisho & 35 & 10 & 7 & 2 & 0 & 3 & 1 & 2 \\
-Oshisho-sama & 36 & 9 & 8 & 0 & 2 & 0 & 2 & 11 \\
\bottomrule
\end{tabular}
\caption{Model Ge15p Keyword Search Results}
\label{tab:model-ge15p-keywords}
\end{table}

\end{document}